\documentclass[10pt,english]{article}

\usepackage{datetime}
\newdate{date}{04}{04}{2020}
\date{\displaydate{date}}

\usepackage{amsmath}
\usepackage{mathptmx}
\usepackage{helvet}

\usepackage{newtxmath}

\usepackage[T1]{fontenc}
\usepackage[latin9]{inputenc}
\usepackage[a4paper]{geometry}
\usepackage{mathtools}
\usepackage{graphicx}
\usepackage{setspace}
\usepackage{esint}
\usepackage[authoryear]{natbib}
\makeatletter
\usepackage{newtxmath}

\usepackage[authoryear]{natbib}\setcitestyle{round}
\bibpunct{(}{)}{,}{a}{,}{,}

\makeatother

\usepackage{babel}

\begin{document}

\title{A Theory of 'Auction as a Search' in Speculative Markets}
\author{Sudhanshu Pani}

\maketitle

\begin{abstract}
The tatonnement process in high frequency order driven markets is
modeled as a search by buyers for sellers and vice-versa. We propose
a total order book model, comprising limit orders and latent orders,
in the absence of a market maker. A zero intelligence approach of
agents is employed using a diffusion-drift-reaction model, to explain
the trading through continuous auctions (price and volume). The search
(levy or brownian) for transaction price is the primary diffusion
mechanism with other behavioural dynamics in the model inspired from
foraging, chemotaxis and robotic search. Analytic and asymptotic analysis
is provided for several scenarios and examples. Numerical simulation
of the model extends our understanding of the relative performance
between brownian, superdiffusive and ballistic search in the model.

\end{abstract}

MODERN MARKETS are built on the foundation of low latency trading and exchange infrastructure. Data available from the order driven markets today can help us to investigate the markets both in high frequency and high resolution. \citet{ohar15} called for the market microstructure toolbox to be enhanced in order to investigate 'orders' that were the unit of information in today's markets. Several recent papers implement objectives that require a high resolution analysis, including, modelling imperfect competition and quote cancellation (\citet{glos19}), price discovery in high resolution (\citet{hasb19}), price discovery in quotes (\citet{brog19}), trading in absence of designated market maker (\citet{kyle18}). In the context of high resolution analysis, the assumption of equilibrium cannot be easily justified. However, such an assumption is fundamental to our current framework used to understand continuous double auctions in financial markets. In walrasian auctions in these markets, the price of a good is defined as the point where the supply and demand curves intersect (a market clearing equilibrium under perfect competition). The relationship between supply and demand and their linkage to price discovery may not however be effectively described by the assumption of walrasian equilibrium in high resolution of analysis. 

Even in quote driven markets, \citet{beja77} had shown that the instantaneous walrasian adjustment imposes prohibitive demands on communication and computation, while the tatonnement iterations must take time, and thus it typically never converges. They further show that market making was necessary to price adjustments away from equilibrium. Modern order driven markets clear in the time scales that ranges from sub seconds to nano seconds. \citet{donier_walras_2016}  report the time required for walrasian adjustments to be in the scale of an hour. \citet{juss19} in their study based on Euronext exchange data and their model (presence of market maker and price discovery criterion) compute the optimal auction duration according to their criterion for 77 European stocks traded on Euronext. They report that the suggested durations are much larger than a few milliseconds, rather of order of 1 to 5 minutes. They reconfirm that continuous limit order book (CLOB) are in terms of their metric sub-optimal. However, the quality of the price formation process in CLOB market is not very far from that of the auction with optimal duration. 

 This paper explores 'Auction as a search' as an alternative to the supply and demand paradigm
for tatonnement in order driven markets under the absence of an auctioneer or designated market maker. The questions we ask include: What is the modern day 'tatonnement' in the electronic order driven markets. Can we explain the dynamics of such markets as a 'search' by buyers for sellers and vice-versa.

The walrasian tatonnement is in effect an algorithm wherein an invisible auctioneer has all information regarding supply and demand schedules and comes up with the price that optimises the quantity traded. This emphasises an important function of the market, that of optimising the traded quantity. Any alternative algorithm needs to meet these two goals. We show in this paper that the tatonnement in continuous double auctions in modern asset markets under the usual price-time priority may indeed be described as a search of buyers by sellers and vice-versa. 'Auction as a search' meets both the above stated objectives. An acceptance of this proposed alternative should help researchers, working on price discovery and trading models in high resolution, build models that are closer to reality.

\section{Introduction}
\subsection{Continuous Double Auctions}
The predominant market design used in asset markets around the world today is the continuous double auction mechanism (although these markets can vary in terms of the trading rules or matching rules they employ). This implies that an asset can theoretically be traded at any point in time. Double auctions enable all participants to provide quotes (both buy and/or sell orders). Futher these markets are organised as electronic exchanges, where the quotes from the traders are matched by matching engines without any official role for intermediaries. The matching is done everytime a new order comes into the market. Typically a mismatch may exist between buyers and sellers at any given instant. Hence, the need for an order based market with two fundamental kinds of orders. Traditionally, it was believed that impatient traders
submited  requests to buy or sell a given number of shares immediately at the best available
price. Such orders are known  as market orders. More patient traders submit limit orders, which also state a limit price, that correspond to the worst allowable price for the transaction. Limit orders may not result in an immediate transaction, and are stored in a queue called the continuous limit order book. Buy limit orders are termed bids, and sell limit orders offers or asks. Both limit orders and market orders play an important role in the order book. Institutional investors, large traders executing meta-orders and algorithmic traders deploy both limit orders and market orders to execute their trades.

A number of approaches have been used to model limit order markets. These include perfect rationality approaches, zero intelligence models and agent based models (refer \citet{gould_limit_2013}). Zero-intelligence models, like the framework introduced by \citet{bak_price_1997}, are useful to model the market activities as stochastic processes.
Researchers have been able to reproduce a number of empirical regularities in the LOB in a zero-intelligence framework using random walk diffusion models. 

Given the continuous and sometimes sequential nature of the auctions, the walrasian model using supply and demand is the accepted solution to describe the equilibrium. Our contention in this paper is that when considering an analyis in high resolution we need to accept the existence of disequilibrium and look for an alternate mechanism for the emergence of prices and  quantity trading.

\subsection{Random Walk to Levy Walk}

The concept of a random walk is a fundamental concept in finance used
to model the stochastic evolution of the price of an asset. If $Y$
$(Y\in R,-\infty<Y<\infty)$ is the log-price of an asset, $Y(t)$
represents the price at time $t(0\leq t\leq\infty)$. The variance
$\sigma^{2}(t)=\langle(Y(t)-\langle Y(t)\rangle)^{2}\rangle$ of the price series explains the dynamics of the asset price. (The brackets $\langle\rangle$ denotes the mean. $\sigma$ is also known as Mean squared displacement (MSD) in diffusion literature).
In Brownian motion, the variance varies linearly with time, $\sigma^{2}\propto(t)$.
Prices in well functioning markets in the absence of arbitrage opportunities,
are known to display brownian characteristics. A non linear
time dependence of the variance may also be observed in the evolution
of asset prices. Non linear time dependence is observed in several
other complex systems. Such a non linear time dependence is a characteristic
of anomalous diffusion, where $\sigma^{2}(t)\propto(t)^{\alpha}$,
with $\alpha\neq1$ (\citet{denisov_levy_2012}). Anomalous diffusion
has been observed in processes with long time correlations (such as
the evolution of the transaction prices or the quoted prices for an
asset). When $\alpha>1$, the case of superdiffusive limit, the random
walker searches his environment much faster than 'Brownian search',
provided the velocity is constrained to a limit so that he remains
within the search space. The levy walk process is a simple stochastic
model that combines these two notions -- a superdiffusive evolution
and finiteness of the velocity of motion. 

The definition of the levy walk model is close to the random walk.
A walker chooses a random direction and a random time $\tau$ and
walks with a constant speed $v$ in the selected direction. After
the time has elapsed a new random direction and a new random time
are picked and the process repeats. What characterises levy walks
is that, the duration of the walks are distributed according to a
power-law density (probability density function, pdf): $\psi(\tau)\propto(\frac{\tau}{\tau_{0}})^{-\gamma-1}$,
where a constant $\tau_{0}$ sets the characteristic time scale and
the exponent $\gamma$ determines explicitly the scaling of the corresponding
MSD, namely $\alpha=1$ when $\gamma>2$ (normal diffusion), $\alpha=3-\gamma$
when $1<\gamma<2$ (superdiffusion), and the choice of the exponent
from the interval $0<\gamma<1$ leads to the ballistic diffusion,
$\alpha=2$. 

The levy walk formalism has been successfully applied in diverse problems
such as the description of DNA nucleotide patterns, modeling the dynamics
of an ion placed into an optical lattice, analysis of the evolution
of magnetic holes in ferrofluids and of photon statistics of blinking
nanodots, engineering of levy glasses, an atom in an optical lattice,
a tracer in a turbulent flow, T- cell motility in the brain , a predator
hunting for food, or a mussel among a bunch of peers. (\citet{denisov_levy_2012},
\citet{zaburdaev_levy_2015}, \citet{reynolds_current_2018}). \citet{raposo_levy_2009}
discuss the developments in random search process that includes levy
formalism and random walks.

\subsection{Levy search in robotics and biology}

Levy search has been observed in biology (foraging and chemotaxis
and others) and used extensively in robotics (a relatively nascent
area of research). Foraging (the search for food) strategies of organisms
have been used for thousands of years and optimised in this time period.
Many of these search strategies (including chemotaxis) involve limited
availability of information. The quest for efficient search algorithms
has been influenced by behavioral science and ecology, where researchers
try to identify the strategies used by living organisms. These provide
motivation for our representation. In biology, levy walk is not confined
to animal foraging and search. Movement patterns resembling levy walks
have been observed at scales ranging from the microscopic to the ecological.
They have been seen in the molecular machinery operating within cells
during intracellular trafficking, in the movement patterns of T cells
within the brain, in DNA, in some molluscs, insects, fish, birds and
mammals, in the airborne flights of spores and seeds, and in the collective
movements of some animal groups. Levy walks are also evident in trace
fossils (ichnofossils) -- the preserved form of tracks made by organisms
that occupied ancient sea beds about 252-66 million years ago. And
they are utilised by algae that originated around two billion years
ago, and still exist today. (\citet{reynolds_current_2018} and references
therein). Existence of levy walk transport predate their formulation
by researchers in the last thirty years. It provides motivation to
model zero-intelligence search for survival using levy walk.

Autonomous mobile robots are required to explore the environment and
locate a target. Targets could range from sources of chemical contamination
to people needing assistance in a disaster area. Levy walks have recently
emerged as universal search strategy in robotics. The finite velocity
in levy walks that restricts search to the search region is useful
in the context of robotics. The foraging trajectories of organisms
look like periods of localized diffusive-like search activity altered
with ballistic relocation to a new spot. This intuition is being explored
in robotics. Performance evaluation of the robot or the search efficiency
is another area where we draw ideas.

Our interest in levy walks as a search strategy emerges from the following:
the speed of the search, the possibility to restrict the search space
and the ability to work with limited information. These are the equipment
that can work in high frequency double auction. We deploy a specific
model of levy walks, ones where the velocities too are a random variable
along with the flight time of the particles. This enables us to represent
the behavioural dynamics of both the return expectations and time
spent in the market by the traders.

In section 2, we propose in our model 'search' as a possible modern
day tatonnement. The model deals with the total order book and is
explained as a diffusion-advection-reaction process in Section 2.2 that also includes effects of imbalance between supply and demand. The diffusion part of the process is the focus
in section 2.1 where in there is no imbalance between supply and demand. Diffusion is considered to be the main mechanism that can explain the price at which auctions are concluded.
Several canonical models in finance have explained the evolution of
price using Reaction-diffusion models and some of them extend to add jumps to
the models (Brownian Semi-martingale or Brownian Semi-martingale jump
process). When volume imbalances exist between buyers and sellers
in the market, biases develop in the system and drift processes can
emerge as additional dynamics to explain the price evolution and volume
trading. Such an imbalance could be general imbalance or point imbalances
at specific prices. 

\section{SEARCH AS TATONNEMENT - The Total Order Book Model}

\subsection{The Search for trades in absence of
bias}

We describe a model where buyers (buy orders) search for sellers (sell
orders) and sellers search for buyers. This is a unique way of looking
at predator-prey relationships (as either the buyer or seller could
take the role of the predator or prey). Only when there is a bias
in the system and an imbalance favouring either the buyer or seller,
there will emerge a true predator-prey situation. We consider the
total order book, comprising the visible part (the limit order book,
LOB) and the hidden part (limit orders and market orders) that will
be placed in the market in due course. The hidden part of the order
book is also referred to as latent order book. We start with an LOB model
based on the framework introduced by \citet{bak_price_1997}, extend
it to the total order book, TOB and later make appropriate changes
to include behavioral dynamics of the agents. Orders are particles
on a one dimensional lattice and their location corresponds to price
(refer Fig 1). When orders representing a buy and sell occupy the
same position on the one dimensional pricing grid it leads to a trade resulting into
a reaction $A+B\rightarrow\emptyset$. 

\subsubsection{The Agents}

We consider two types of agents in our model. The first can be characterised
as noise traders with zero intelligence. The noise traders start their
journey by placing market orders at the last transaction price. However,
in the model they are not required to absorb losses. They are zero
intelligent because once they transact through a market buy (sell)
order, they place a subsequent limit sell (buy) order to occupy the
price at the best offer (bid). In the simulation of the model (Section
3), for example, we implement this by placing a limit offer at 0.1
USD (an arbitrary choice) over the transaction price they record as a market order or a
limit buy at 0.1 USD lower than the transaction price. Once such a
limit order is transacted they can again go back to place market order
and the cycle repeats. The endevour is to remain in the market and
get involved in as many transactions.

The second type of agents are the strategic traders. These agents
represent a heterogenous group of investors who have a view on future
transaction prices. Or they may have specific portfolio execution
mandates. They seek specific prices to enter into transactions. The
origin of pricing decision (as well as quoted volume) may be varied: 
portfolio decisions, private information, liquidity concerns, meta-order
execution, fundamental value etc. Strategic traders may be buyers
or sellers and use limit orders or market orders to execute their
transactions. When they place market orders, they come into the market
at the specific price and expect execution of the trade. They leave
the market after an execution.

\subsubsection*{Order cancellation by agents}

Cancellation of orders sets up useful behavioural dynamics in these
markets. \citet{benzaquen2018fractional} use cancellations to model
the supply and demand in the order book. In section 2.1, neither the noise
traders nor the strategic traders come back to the market in
case orders are cancelled before execution. Quotes from noise traders
are cancelled in section 2.1 and 2.2 under two circumstances. First,
when the particles end their designated flight time and second in
case the available budget with the trader falls short of a limit ask
order. We treat it as a default in the trading process and remove
the quote and trader. Quotes from strategic traders are cancelled
in both section 2.1 and 2.2 when they end their flight time. Additionally,
in the examples in section 2.2 to illustrate the traders response
to biased trading conditions, we allow orders from strategic traders
to be cancelled and a new order placed again. These are quote revisions.

 The merit of our model is to show that in high resolution price discovery and trading of volume
in financial markets can be achieved using a model of search.

\begin{figure}
\includegraphics[scale=0.8]{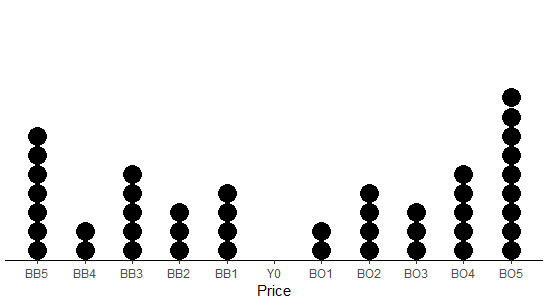}

\caption{One dimensional lattice showing price and orders shown as particles
stacked up at the price. If a buy and sell particle occupy a single
lattice point it leads to a transaction. Y0 is the last transaction
price and does not have any market particles. }
\end{figure}

Consider a continuous auction, order driven market, in a relatively
homogeneous state. A homogeneous state in the order book can be achieved
by a constant gradient or uniform gradient in the quantity density.
The equivalent in economics perspective would be that the demand and
supply is balanced. We want to avoid perturbations due to larger buy
or sell orders causing bias in the system. We envisage a model where
market buy (or sell) orders search for targets - sell (or buy) orders,
to effect a trade. The target could be a market order or limit order.
A market order is one where a transaction can get effected immediately
if a counterparty is available. The behaviour of the agents represented
in the form of orders is treated as particles. One may consider a
single particle to be equivalent to the smallest quantity of the order
that is allowed to be placed in the market.

\subsubsection*{The Total Order Book}

The total order book comprises the limit order book (visible LOB)
and the latent orders. Latent orders are orders that are yet to be
placed in the market, but the traders have already planned the same.
These include the limit orders that would be placed in future and
the market orders. Our conceptualisation is not based on price but
on the quoting strategy of the agents. Thus when orders are present
in the latent order book, the traders have decided on the price, time,
sequence or events at which they would introduce the orders into the
visible order book. A market order in the latent (hidden) order book
is introduced as a market order necessarily at specific predecided
price, time or sequence. The order does not carry a price, but is
only introduced when the transaction price is around the predecided
price. A limit order in the latent book, in contrast, will carry a
limit price and would be introduced by the traders into the visible
LOB at a predecided time, sequence or event. Such a representation
addresses the practice of algorithmic trading and slicing of metaorders
into child orders.

Fig 2 represents the total order book. It super imposes the strategic
and noise particles into the limit order book model depicted in fig
1.

\begin{figure}
\includegraphics[width=1\columnwidth]{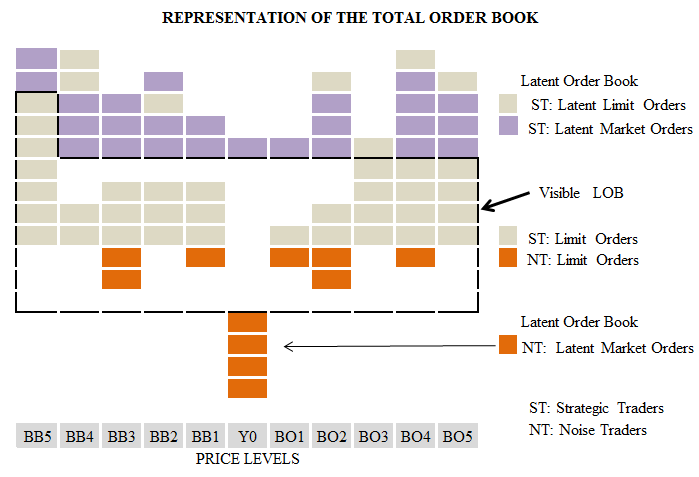}

\caption{Representation of the Total Order Book as a one dimensional lattice.
It comprises the visible limit order book and the latent order book.
The dark border represents the visible limit order book given in fig
1. The limit orders from strategic traders (ST) are in grey. The limit
orders from noise traders (NT) are placed immediately after a market
order is executed by a NT. Strategic traders can place limit orders
(grey) or market orders (purple) in future and hold their orders in
the latent order book. Noise traders place market orders at the last
transaction price Y0. BO-Best Offer; BB- Best Bid.}
\end{figure}

To investigate the tatonnement process, we note that any transaction
necessarily requires a market order as atleast one of the counter
party. We suggest that modelling the market orders should be able
to explain all transactions and help us achieve our objective. The
reduction in the dimensionality is an important benefit.

Let $u(x,t)$ be the probability density function (pdf) of the distance
travelled by market order particles in search of trades. Thus, $x$
is the total length of the jumps and $t$ is the total time. We can represent $u$ as a continuous time random walk. Let, $Y_{0}$  be the last traded price and $Y_{1}$ be the next expected transaction
price. When $Y_{0}$ occurs, it ends the flight of some of the market
order particles (those that have reached or crossed the price $Y_{0}$).
However, some particles are still in flight. Between, $Y_{0}$ and
$Y_{1}$, a number of new quotes are placed into the book, some particles
complete their flights through cancellations and at least one would
complete their flights with the transaction ending at $Y_{1}$. We
consider particles that have completed their search with success or
ended with cancellation. (refer Fig. 3 where $Y_{0}=Y0$ and $Y_{1}=BO1$). 

We first specify the price quoting and cancellation behaviour and then we
shall resume with the model. 

\begin{figure}
\includegraphics[width=1\columnwidth]{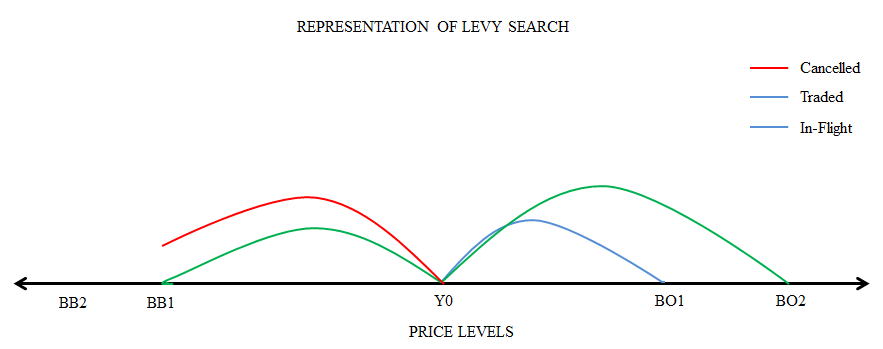}

\caption{Levy Search. The fate of 4 market orders particles introduced at the
last transaction price Y0 is represented. These particles are introduced
sequentially starting time $t_{0}$ after the last transaction occurs.
The particles differ in the velocity imparted to them and time spent
in the market before cancellation. Both these variables are random
variables in the model. The sign of the velocity gives the direction
of flight and represents a buy order or a sell order. The market orders
can be placed by both noise traders and strategic traders. The next
transaction occurs when the first particle reaches the target at time
$t_{1}$, in this example, the best offer BO1, which is assumed to
have a limit order. At this instant three particles are still in flight.
Two market particles reach targets at BB1 (time $t_{2}$) and BO2
(time $t_{3}$)and are transacted sequentially. One particle is cancelled
before it can reach the target. Between time $t_{0}$ and $t_{3}$many
more orders may have been introduced and it is not represented in
the figure.}
\end{figure}

\paragraph*{Quoted Price and Cancellation}

Traders display their intentions by placing market and limit orders.
They impart a velocity $v$ to the particles and the orders have a
flight time $\tau$. The flight ends with a quote cancellation or
trade and the particles are believed to be at rest before entering
the market. The particles enter the market either into the LOB or
into the latent order book as limit or market orders . The length
of the walk or jump of a particle can be represented by the coupling
$v\tau$. This is close to the model in \citet{zaburdaev_random_2008}
that dealt with a random walk with random velocities. 

The velocity imparted to the orders from strategic traders depends
on certain dynamics experienced by the traders and is an important
element that determines the quoted price. For any trader, it is governed
by his expectation of success. The strategic traders have specific
prices that they want to quote. This price may also be viewed relative
to the return they desire from the first transaction price of the
trading day. Let the expectation of this return for the $jth$ trader
be $\langle r_{j}\rangle$ and the averages of the velocity and flight
time be denoted by $\bar{v_{j}}$ and $\bar{\tau_{j}}$ respectively.
So, the following holds:

\begin{equation}
\langle r_{j}\rangle=\bar{v_{j}}\bar{\tau_{j}}
\end{equation}

The above dynamics is guided by the past experience of success, current
trading conditions and state of the order book. We do not investigate
these dynamics and to simplify, we treat the velocity and flight times
as random variables drawn from distribution. The velocity imparted
along with the flight time determines the quoted price. The strategic
traders move out of the market after a trade execution or at the end
of the flight time.

Noise traders quote the last transaction price in their market orders
and a small profit over the last transaction price while quoting limit
orders. This over-rules any role of velocity. The flight time in case
of quotes from noise traders determines when any quote from the noise
trader is cancelled and the trader moves out of the market.

The particle is assumed to be in-flight if its current price (last
transaction price after which it entered the market + the jump) is
still less than the transaction price quoted and it is not cancelled.
The velocity $v$ can have positive and negative values to include
buy or sell orders (direction of motion). The inter trade time period
is known as a trade duration. If $'N'$ number of trades takes place
in a unit of time $dt$, the average duration can be given by $\langle\mu\rangle=dt/N$.
Each inter-trade period is a search for the next transaction. A number
of jumps get recorded in this period. 

Equation (2) gives the total path traversed in the search $x$ till time $t$. Driven by return expectation in (1) the quote from a trader is provided with a velocity drawn from a pdf $h(v)$ and the flight time (cancellation behaviour) from a pdf $f(\tau)$. The distance travelled by such a particle during search is given by the coupling $v\tau$. 

 Flight time and velocity are the two basic and independent random variables
of the model. They are normalised to 1, $\int_{-\infty}^{\infty}h(v)dv=1$
and $\int_{0}^{\infty}f(\tau)d\tau=1$. In the base case, to avoid
a bias in the system, the velocity distribution is symmetric, $h(v)=h(-v)$.
However, interactions due to limit orders or transactions in the system
can change this assumption and we shall model this in section 2.2.

\begin{equation}
u(x,t)=\int_{-\infty}^{\infty}dv\int_{0}^{t}u(x-v\tau,t-\tau)h(v)f(\tau)d\tau+n_{0}(x)\delta(t)
\end{equation}

The right side in (2) describes the dynamics of  search where the distance of $v\tau$ is traveresed in time $\tau$ so that the total distance traversed in search reaches $x$ from $x-v\tau$ and total time spent in search reaches $t$ from $t-\tau$. Taking into account the possible velocity-time coupling, the joint probability $h(v)f(\tau)$, the first term in (2) integrates over all flight time and velocity combinations. Since a coupling of velocity and time represents an order, it in effect integrates the distance travelled for all search orders. For particles that have completed their flights, they would have
reached a point on the lattice grid. It may or may not have resulted
in a success. For all such flight times, we sum the total distance travelled
or jumps for all velocities. Particles still in flight are not included. The initial distribution of the market particles is given by $n_{0}(x)$. This search is the tatonnement in modern markets. 

The success of each search is a trade transaction. The pdf of the
traded particles, $q(x,t)$ is simply the number of trades that take
place in a time interval $(0,t)$. However, the trades being the success
of the search of market particles in a tatonnement, we define the
efficiency $\eta$ of the search process as the ratio of the total
path traversed in the search to the number of targets searched or
trades done. This gives us a link to represent $q(x,t)$ in terms
of $u(x,t)$ and $\eta$ as in equation (3). $\eta$ is again a function
of the coupling of velocity imparted and the flight times. 

\begin{equation}
q(x,t)=\int_{-\infty}^{\infty}dv\int_{0}^{t}\frac{u(x-v\tau,t-\tau)h(v)f(\tau)}{\eta(v\tau)}d\tau
\end{equation}

Equations (2) and (3) fully describe the dynamics of the system with
a given initial density of particles and the two pdf for the flight
times and velocities. They establish the crucial link between the
tatonnement in the auction with the trades. Next we solve these equations
analytically. First, we determine the total path traversed by market
particles in the search process and then introduce the result into
the equation of the trade density. We apply the Fourier transform
with respect to the spatial coordinate in (2) and then the Laplace
transform with respect to time. This yields in (4) the path traversed
in search in the Fourier-Laplace domain, $k,s$, as:

\begin{equation}
u_{k,s}=\frac{n_{0,k}}{1-[h_{k\tau}f(\tau)]_{s}}
\end{equation}

We can introduce the result in (4) into the Fourier-Laplace expression
for equation (3) to get the analytic expression for the density of
trades in the Fourier-Laplace domain, obtained through the search
process in (5).

\begin{equation}
q_{k,s}=\frac{n_{0,k}}{\eta_{k,s}}\frac{[h_{k\tau}f(\tau)]_{s}}{(1-[h_{k\tau}f(\tau)]_{s})}
\end{equation}

To find a solution, the next step is to take the Laplace inverse of
equation (5). However, an analytic representation and direct inversion
of the equation (5) is not feasible. \citet{froemberg_asymptotic_2015}
recommend an asymptotic analysis for large space and time scales,
$x,t\rightarrow\infty$. We use the same approach.
Going to Fourier-laplace space using the tauberian theorem, this limit
corresponds to $(k,s)\rightarrow(0,0)$ such that
$k/s=constant$. This has to be performed numerically.

It is important to define the velocity and flight time distributions
before we move to obtain the inverse transformation. In our view this
needs to come from empirical analysis. Further, velocity distribution
cannot be obtained directly as it is a notional quantity and needs
to be interpreted from equation $(1)$ that describes the relationship
with returns.

As discussed in \citet{zaburdaev_random_2008}, \citet{froemberg_asymptotic_2015}
and \citet{zaburdaev_levy_2015} when the velocity distribution is
Cauchy or lorentian, the density of the particles also is lorentian,
independent of flight times and jump lengths. Such a lorentian velocity
profile appears in real physical phenomena such as two dimensional
turbulence and is also found in model distributions of kinetic theory,
statistics, plasma physics and starving amoeba cells. We know that
a cautchy process does not give rise to a continuous sample path for
the price and it differs from Brownian motion as there are large jumps
not infrequently. As given below, we make arbitrary choice of a lorentian
velocity distribution $h(v)$ and intuitively a flight time distribution
$f(\tau)$ with power tails.
\begin{equation}
h(v)=\frac{1}{u_{0}\pi}\frac{1}{(1+(\frac{v^{2}}{u_{0}^{2}}))}
\end{equation}

\begin{equation}
f(\tau)=\frac{\gamma}{(1+\tau)^{1+\gamma}}
\end{equation}

In equation (6) $u_{0}$ is needed to constrain the velocities so
that the particles do not go beyond the ballistic cones, else it will
lead to instantaneous dispersion. The $\gamma$ in equation (7) is
varied to get different transport. In equation (5), given the asymptotic
limit we want to evaluate, we further set a constant efficiency, so
that there is an expression for the initial density of particles for
the trade density. The propagator for our model can then be noted
as follows:

\begin{equation}
G(k,s)=\frac{\mathcal{L}[f(\tau)h(k\tau)]}{1-\mathcal{L}[f(\tau)h(k\tau)]}
\end{equation}

where $k\tau$ is the Fourier variable conjugate to $v$. The equation
(8) retains the form of the well known Montroll-Weiss equation for
the pdf of the uncoupled continuous time random walk (CTRW) to find
the particle $x$ at the time $t$, modified such that it applies
to random jumps in velocity. Equation (8) can be rewritten as (see
Appendix C for details):
\begin{equation}
G(k,s)=\frac{\intop_{-\infty}^{\infty}dvf(s+ikv\tau)h(v)]}{1-\intop_{-\infty}^{\infty}dvf(s+ikv\tau)h(v)]}
\end{equation}

For the flight time distribution we have chosen and in the long time
limit the expansion in the Laplace space is given by,
\begin{equation}
f(\tau)\approxeq1-\tau^{\gamma}\Gamma(1-\gamma)s^{\gamma}
\end{equation}

Using (10) the asymptotic version of (9) is,

\begin{equation}
G(k,s)=\frac{1}{s}\frac{\intop_{-\infty}^{\infty}(1+ikv/s)^{\gamma-1}h(v)dv}{\intop_{-\infty}^{\infty}(1+ikv/s)^{\gamma}h(v)dv}
\end{equation}

\subsubsection*{Random walk with Ballistic scaling}

As the simulation in section 3 confirms, ballistic scaling may be
useful in thinly traded stocks where the arrival rates of orders,
transactions and cancellations is low. A method exists for the inversion
of the fourier laplace expression for propagators with ballistic scaling.
A propagator of a random walk model has ballistic scaling if it can
be written in the form: $G(x,t)\approxeq\frac{1}{t}\phi(\frac{x}{t}),\,t\rightarrow\infty$,where
$\phi$ is the scaling function. In Fourier-laplace space this is,
$G(k,s)\approxeq\frac{1}{s}g(\frac{ik}{s})$. Comparing the two forms
we can rewrite the scaling form of our equation as in equation (12),
where $\xi=\frac{ik}{s}$ . 
\begin{equation}
g(\xi)=\frac{\intop_{-\infty}^{\infty}(1+\xi v)^{\gamma-1}h(v)dv}{\intop_{-\infty}^{\infty}(1+\xi v)^{\gamma}h(v)dv}
\end{equation}

Using a arbitrarily chosen special situation induced by Cauchy distributed
velocity and taking the inverse Laplace and Fourier transform we get
a form of cauchy distribution as in (13) (refer Appendix B). 

\begin{equation}
G(x,t)=\frac{u_{0}t}{\pi(u_{0}^{2}t^{2}+x^{2})}
\end{equation}

Although $u_{0}$ comes in due to the need to restrict the particles
in the ballistic cone, it is an important feature as it slows down
the diffusion process. The density of trades scales inversely with
time and square of the price.

\subsubsection*{Role of diffusive search and the traders}

The role of diffusive levy search is to provide continuity in the
auction and also volume trading. The nature of this continuity can
be evaluated basis the success of the search in terms of the density
of trades. This is further related to the inter trade duration. The
parameters that determine the trade density are thus critical to continuous
auction. First, the presence of noise traders and the market orders placed by noise traders and strategic traders. Market orders need to be always available to provide
continuity in auctions. Else, traders need to cancel and revise their
quotes, which we shall discuss in section 2.2. Trading can slow down
if less number of noise traders are present or if they get locked
away post transaction either in limit orders or leave the market. We always expect strategic traders to be present in the market. But presence of noise traders can be impacted by market design.
 Second, the efficiency of the search. The efficiency is dependent
on the resource density or target density. This comprises both of
the limit orders and strategic market orders that come in whenever
a new price is discovered. It is these market orders that bring in
efficiency improvement by leading to greater number of transactions
at already discovered price. 

\subsubsection*{Efficiency of Search}

A useful tool to judge how a trading system is performing and in the
case of this model, how the search for trades performs is to
evaluate a global efficiency of the search using the trade duration.
The trade durations is the time diference between one trade and the
next. Thus it is the time between any two auctions in continuous trading
set up. The reciprocal of the trade duration is rate of trades. If
the searchers are taking longer flight times or longer distance in
the walk before they locate the target, the rate of trades would decrease.
This measure is an indirect route to use the concepts like distance
travelled per success or average flight time per success, but is more
meaningful and easier to evaluate for a trading system. The rate of
trades or efficiency of search can be computed as $\langle\frac{1}{\tau}\rangle,$where
$\tau$ is the trade duration. A similar measure for efficiency of search
in a different context was also used by \citet{palyulina2014levy}.
In section 3 (numerical simulation) we have used this measure (rate
of trades) as the efficiency of search to compare search in ballistic,
superdiffusive and brownian regimes with different market conditions.

\subsubsection*{Need for velocity as random variable}

The use of Random velocities in the modelling is unique in the context
of modelling stock prices. It is not required when modelling the trade
prices, but required when modelling the quotes. As discussed earlier,
the velocities are a notional quantity and cannot be directly observed.
This technique is a useful tool to connect the random walks of quotes
to the trade price series. Most quotes are cancelled sooner or later
and hence do not get involved in the reaction. The random velocities
through the return expectations can connect the trade durations to
the quote durations. 

\subsection{The Complete Model - Search for trades in presence of bias}

In section 2.1 we dealt with a limit order market (Total Order Book)
in homogeneous state where the search is diffusive. A situation where
there is no shortage of noise traders and strategic traders have placed
market orders at widely dispersed prices. The diffusive search was
crucial to explain price continuity. This section combines diffusive
search with other mechanism that can explain auctions when we relax
the assumption of 'no bias'. In case of excess demand (buy or sell
side) of liquidity a bias will get generated in the system, resulting
in a drift. The market will move to improve the search efficiency
due to higher availability of resource targets, i.e more and more
targets can be acquired at the existing transaction price. The number
of targets thus go up without increase in the path travelled in the
search leading to higher efficiency. The market will naturally drift
from resource rich regions to regions where market order particles
may be available. This is because market orders from strategic traders are only placed when the relevant prices and time is reached. 

Further, there is a stochastic evolution of the
transaction price of the asset and quantities traded. This is the
result of the strategic games between the market orders and limit
orders, buyers and sellers, new order arrivals, order cancellations
and transactions. We allow for such a stochastic evolution although
we model the dynamics of market order particles only. Since market
order are involved with each transaction, we expect to successfully
explain the stochastic evolution of the transaction prices and volume
traded. \citet{donier_fully_2015} use a diffusion-advection-reaction
model to explain the evolution of the marginal supply and demand in
the market. We set up a model where market order particles search
for targets in the double auction market. While we classify our model
as a diffusion-advection-reaction model, given the view point of search,
it has been built using concepts involving foraging, predator-prey
search and chemotaxis (\citet{grunbaum_using_1998} and references
there in). 

In section 2.1, $Y(Y\in R^{+},0<Y<\infty)$ represented the price
of an asset and $Y(t)$ the price at time $t(0\leq t\leq\infty)$.
We denote the price by $x$ in this section (in section 2.1 $x$ denoted total distance travelled). 

Let $g(x,t)$ represent
the density of market particles at a price $x(x\in R^{+},0<x<\infty)$
and time $t(0\leq t\leq\infty)$. Our fundamental set up is given
by the partial differential equation (14).

\begin{equation}
\frac{\partial g(x,t)}{\partial t}=
\end{equation}
\[
\frac{\partial }{\partial x} (D(t)\frac{\partial g(x,t)}{\partial x}-\lambda_{1}(t)\frac{\partial f(x,t)}{\partial x}g(x,t)-\lambda_{2}(t)g(x,t))+v(t)g(x,t)+q(x,t)
\]

\subsubsection*{The set up in equation (14) }

(14) explains the stochastic time evolution of the density of market
order particles in our one dimensional grid. The first term on the
right hand side represents diffusion that we discussed in section
2.1. The second and third terms relate to volume trading and price
evolution due to drift. The fourth term is the net addition of market
particles. The last term represents large external additions of market
particles into the system.

Section 2.1 can explain a walrasian tatonnement where prices are set
when the quantity demanded by buyers matches the quantity supplied
by sellers. Modern order driven markets operate at a speed where there
is neither sufficient time to establish a walrasian equilibrium nor
there is an official market maker who will adjust the price instantaneously.
Hence, transactions take place at prices out of equilibrium. Traders
have always observed markets operating out of equilibrium. Even in
Quote driven markets as noted by \citet{beja77} and \citet{beja1980dynamic},
set up of equilibrium needs more time to allow for the information
flow required for the purpose. We look at (14) as the mechanism that
can take the trading system towards equilibrium. We now explain how the
terms enter the right hand side in (14) :
\begin{itemize}
\item Diffusion (first term): This is the primary search mechanism through which price
continuity and quantity transport takes place. In the absence of bias
this will remain the only mechanism (already discussed in the Section 2.1). 
\item Drift due to target density (second term): We implement a formulation widely used
in foraging studies ( \citet{grunbaum_using_1998}). In the set up
of those studies, the drift velocity is proportional to the gradient
resource density. And the coefficient is the taxis coefficient. The
use of resource density implies the dimensionality
of the model increases. We, however, note that the density of targets (limit
orders or market orders) in our use case should result either into
a response or consequence in the system. We model the drift resulting
from this response or consequence in market particles. $f(x,t)$ is
the perceived resource density basis the actions in the system. In
this innovation we conjecture that the drift resulting from a gradient
of resource density should have a mirror image in the searcher. This
mirror image may be different than the original gradient, but
is more relevant as it is likely to be a tactical response in the
absence of complete information (since the visible order book is not
the complete order book). We may err in magnitude but not direction
and the system will try to iteratively guess the true gradient of
resource density. Again we accomodate a time evolving coefficient
$\lambda_{1}$.
\item Drift due to movement within the market order particles  (third term): There are
several scenarios we envisage where a number of market particles could
get activated into directed activity resulting in a drift. For example,
high transaction activity on one side of the market, will result in
noise traders getting released from limit orders suddenly due to the
reactions. This early release increases returns for noise traders.
If these traders sense the action to continue, they would rush to
the other end to repeat the cycle. Similarly, traders yet to place
orders may anticipate change in resource gradient and resource density
due to news flow and react by adjusting their own positions. $\lambda_{2}$
is a time varying coefficient .
\item Net addition to market particles density $v$  (fourth term) : $v(t)$ is a time
varying composite coefficient of the new order additions $v_{a}$,
cancellations $v_{c}$ and transactions (reactions) $v_{r}$. $v(t)g(x,t)=v_{a}(t)g(x,t)-v_{c}(t)g(x,t)-v_{r}(t)g(x,t)$.
New order additions of market particles includes market particles,
hitherto latent, coming into the order book, new traders, limit orders
that are cancelled and placed as market orders as the traders become
impatient. Hence, this can be a non-trivial component. Cancellations
and transactions remove market particles of strategic traders from
the order book. Noise traders move out of market only at end of their
flight time. 
\item $q$ (fifth term) is a source of market particles much larger than what comes in
during the average trading day for the asset. It could be a demand
or a supply of liquidity. The source may be a single or few point
sources or distributed across a price range and even in time.
\end{itemize}

In summary, we build a comprehensive model of stochastic evolution
of market particles. Expand (14) to obtain (15) :

\begin{equation}
\frac{\partial g(x,t)}{\partial t}=
\end{equation}
\[
D(t)\frac{\partial^{2}g(x,t)}{\partial x^{2}}-\lambda_{1}(t)\frac{\partial f(x,t)}{\partial x}\frac{\partial g(x,t)}{\partial x}-\lambda_{1}(t)\frac{\partial^{2}f(x,t)}{\partial x^{2}}g(x,t)
\]
\[
-\lambda_{2}(t)\frac{\partial g(x,t)}{\partial x}+v(t)g(x,t)+q(x,t)
\]

Equation (15) can be solved analytically. We draw upon the technique
used by \citet{sanskrityayn_analytical_2016}, who used the Greens
function method to solve their diffusion-advection equation in the
context of pollutant solutes in the atmosphere. To solve the equation,
we note that $q$ will remain untouched and we need to reduce the
equation to a known form so that we find an expression for $f(x,t)$.
We do a co-ordinate transformation from the domain $(x,t)$ to the
domain $(X(x,t),t')$. The domain X is essentially fixed time-snapshots
of the entire lattice. We want to transform (15) to the form in equation
(16).

\begin{equation}
\frac{\partial G(X,t')}{\partial t'}=D_{1}(t')\frac{\partial^{2}G(X,t')}{\partial X^{2}}-\lambda(t')\frac{\partial G(X,t')}{\partial X}+v_{1}(t')G(X,t'))+q_{1}(X,t')
\end{equation}

Using the domain transformation, we can write equation (15) as equation
(16), following which we equate the coefficients to solve for the
interim variables and transformations introduced. (Refer Appendix
D for details). Using the transformations we can reduce our initial
equation to the following form,
\begin{equation}
\frac{\partial G(X,t)}{\partial t}=\frac{D(t)}{\beta^{2}(t)}\frac{\partial^{2}G(X,t)}{\partial X^{2}}-\lambda(t)\frac{\partial G(X,t)}{\partial X}+v_{1}G(X,t)+q_{1}(X,t)
\end{equation}

where the dimensionless expressions and transformation are given in
(18), (19) and (20).
\begin{equation}
\frac{D_{1}(t)}{D(t)}=\frac{1}{\beta^{2}(t)}
\end{equation}

\begin{equation}
\beta=e^{\int_{0}^{t}(v(s)-v_{1}(s))ds}
\end{equation}

\begin{equation}
X=\frac{x}{\beta(t)}+\int_{0}^{t}(\lambda(t)-\frac{\lambda_{2}(t)}{\beta(t)}-\lambda_{2}(t)\frac{\phi_{2}(t)}{\beta(t)})
\end{equation}

The initial conditions for this equation are $G(X,0)=G_{i}\omega(X),$
with $-\infty<X<\infty$ and $t>0$ . Next, we try to remove the drift
term and the decay term from (17) by using further transformations.
We use the following transformation equations one after the other
for this purpose. In (22) $\beta_{1}=e^{\int_{0}^{t}v_{1}(s)ds}$is
a dimensionless term and in (23) $\beta$ is as per (19). In (23)
$T$ is a time variable. Equation (17) now reduces to (24).

\begin{equation}
\eta=X-\lambda(t)t
\end{equation}

\begin{equation}
K(\eta,t)=\frac{G(\eta,t)}{\beta_{1}(t)}
\end{equation}
\begin{equation}
T=\int_{0}^{t}\frac{1}{\beta^{2}(s)}ds
\end{equation}

\begin{equation}
\frac{\partial K(\eta,T)}{\partial T}=D\frac{\partial^{2}K(\eta,T)}{\partial\eta^{2}}+\frac{Q(\eta,T)\beta^{2}(T)}{\beta_{1}}
\end{equation}

We now need to solve equation (24) to obtain a master equation for
transport of market particles in double auction limit order asset
markets. \citet{haberman_elementary_1987} provides a solution for
equations such as (24) using Greens Function Method (GFM). The solution
to (24) based on GFM is given in (25):

\begin{equation}
 K(\eta,T)=\int_{0}^{T}\int_{-\infty}^{\infty}\frac{Q(\chi,\tau)\beta^{2}(\zeta)}{\sqrt{4\pi D(T-\zeta)}\beta_{1}}exp(-\frac{(\eta-\chi)^{2}}{4D(T-\zeta)})d\chi d\zeta
\end{equation}

\[
+\int_{-\infty}^{\infty}\frac{1}{\sqrt{4\pi DT}}G_{i}\omega(X)exp(-\frac{(\eta-\chi)^{2}}{4DT})d\chi
\]

Next we sequentially trace back the transformations done earlier,
in reverse order to get the solution below in (26). Here, $\zeta=\int_{0}^{\tau}\frac{1}{\beta^{2}(s)}ds$
and the initial condition $g(x,0)=G_{i}\omega(x)$.

\begin{equation}
g(x,t)=\beta_{1}(t)\int_{0}^{t}\int_{-\infty}^{\infty}\frac{Q(\chi,\tau)}{\sqrt{4\pi D(T-\zeta)}}exp(\frac{(\frac{x}{\beta}-\int_{0}^{t}(\frac{\lambda_{2}}{\beta}+\frac{\lambda_{1}\phi_{2}}{\beta})ds-\chi)^{2}}{4D(T-\zeta)})d\chi d\tau
\end{equation}
\[
+\beta_{1}(t)\int_{-\infty}^{\infty}\frac{1}{\sqrt{4\pi DT}}G_{i}\omega(\chi)exp(-\frac{(\frac{x}{\beta}-\int_{0}^{t}(\frac{\lambda_{2}}{\beta}+\frac{\lambda_{1}\phi_{2}}{\beta})ds-\chi)^{2}}{4DT})d\chi
\]

Before we move into asymptotic analysis, a few observations are due:

\subsubsection*{The significance of $\beta$.}

The dimensionless quantity $\beta$ has a physical meaning and is
not simply an artefact of domain transformation. On transforming the
domain from $(x,t)$ to $(X,t')$, each point on the one dimensional
lattice is now a space-time variable. The advantage of this construction
is that $X$ alone can explain the time evolution of the system. We
have used this intuition to our advantage. $\beta$ relates the diffusion
in the new domain to the original domain in (18). Should the diffusion
coefficient be different in the two domains, since we have defined
diffusion coefficient as a variable only in time. Since, $\beta$
is defined in (19) in terms of the relative contribution from the
net addition to the density of market particles in the two domains,
we infer that at any particular price in the domain $(X,t')$ we are
able to witness the impact of the deposition, cancellation and transaction
over time. Thus the behaviour seems to be that $v_{1}$ already incorporates
time evolution, while $v$ has to be integrated over time. As we shall
see, this makes inference from asymptotic analysis easier.

\subsubsection*{Quantity transport and price auction}

We are intuitively claiming quantity trading. A change in density
of market particles on the grid means a reaction if particles with
a buy and sell intention occupy a grid position in the LOB (not latent
orders). A difference between physical systems and asset markets is
that while in the former ensemble measures are defined on the basis
of averages, in the latter the representative measure of price is
determined by the last transaction, even if it may have involved the
minimum allowed quantity. This coupled with low granularity in minimum
quantity and non-discretisation of the price grid has led to the possibility
of viewing the price auction and quantity auction as separate interconnected
mechanisms. We may see auctions resulting in transactions where prices
remain stable with low quantity trading or a large amount of quantity
traded without price impact or quantity traded with price impact or
price impact with low quantity trading.

\subsubsection*{Drift and diffusion}

The more market order particles diffuse, new trade transaction prices
are formed. But the noise traders can get stuck in longer waiting
times in limit orders. The density of the market particles may not
allow absorption of volume / quantity of resource density. Sometimes,
with targets remaining unconquered the resource particles (limit orders
or market orders) have a tendency to move in opposite direction to
diffusion of market order particles. We have modelled the impact of
this in the drift term that opposes general diffusion. The drift is
dependent on the quote revisions from the traders as discussed in
example in section 2.2.2.

\subsubsection*{Bid-ask bounce}

Another interesting facet is the presence of a reaction front characterised
by bid-ask bounce. The diffusion through levy search has a tendency
to pull the transaction price in opposite direction resulting in the
prices moving from the buy side to the sell side. A counter force
to this would come from the fact that the participating noise traders
would get locked in limit orders and hence the number of transactions
they can enter into decreases thereby gradually reducing the efficiency.
Sell market orders introduced on the ask side of transaction prices,
would need to travel longer to reach the buy side and vice-versa.

We now look into specific scenarios that may operate in the market
through simple examples. Such examples can be generated in conjunction
with a trading model that employs search as a tatonnement as in the
Intraday Trading model of \citet{pani19}.

\subsubsection{A Steady State Solution}

We want to find a steady state solution to our master equation. Say
we have total addition of market particles other than usual as $Q(x,t)$.
A steady state solution is an equilibrium solution when the change
of market particle density over time is zero. It generally refers
to the state where the long term average volume in the asset gets
traded. In steady state the diffusion coefficient and drift velocity
coefficients are not time dependent. The contribution to market particles
coming from $v$ stabilises. In steady state, $v=v_{1}=constant$
i.e the instantaneous change is equal to the total change. At any
time $t_{i}$ the net change is effected only at a particular $x$
and not other locations. When price priority rule is used for matching
trades, it results in such a general condition in the limit order
markets. The rate of particles released from limit orders following
transactions is constant and transaction rate is constant.There are
two ways to find the solution, first by setting $\frac{\partial g}{\partial t}=0$
or second, we find the solution from the master equation (26) when
$t\rightarrow\infty$. 

Our basic equation for the gradient function in the drift velocity
expression is (refer Appendix Eq. D.12): 
\[
-(v(t)-v_{1}(t))\sqrt{\frac{D_{1}(t)}{D(t)}}=\frac{\partial}{\partial t}\sqrt{\frac{D_{1}(t)}{D(t)}}
\]
When $t\rightarrow\infty$. the perceived resource
gradient is constant, i.e $\frac{df}{dx}$ is constant and $\frac{d^{2}f}{dx^{2}}=0$. This implies $v=v_{1}=constant$. What happens after a long time post addition of Q, with sufficient
time to all the three processes to play out to reach an equilibrium
state. As $t\rightarrow\infty$, $v-v_{1}=0$
and $\beta=1$. However, $T$ becomes very large and indeterminate
value. The second term of the equation (26) disappears. $T-\zeta=e$.
$\beta_{1}$ is a constant.

Further, since transaction rates are constant and not high $\lambda_{2}=0$.
There is no drift, hence $\lambda_{1}=0$. The equation reduces to
(27) with the assumption that $\frac{Q(x)}{\sqrt{4De}}=\frac{-(x-\chi)^{2}}{4De}$.
Thus we recover the square root law where $\chi$ is the impact cost.
$x-\chi$ is the price region where we will have trades to resolve
$Q$.

\begin{equation}
g(x)=\frac{\beta_{1}}{\sqrt{\pi}}e^{\frac{-Q(x)}{\sqrt{4De}}}
\end{equation}

\subsubsection{A large demand}

What happens when a large demand $M$ comes into the market. $M$
could be a single large order or a few orders in a short span of time.
If the demand is a new  market order, it would follow a path trying
to move towards a steady state dynamics. Otherwise, after the trades
hit this block, it takes a few transactions to lock up the noise particles
in the next best limit order position. The market order particles
start to sense a gradient and drift begins at a later stage when
the locked noise particles are released.
The following example illustrates this mechanism that we offer for
the market. Fig. 4 to Fig 8 also illustrate some of the discussion
below. 

\subsubsection*{Example}

Let $Y_{t}$ denote the transaction price. The grid is denoted relative
to the first transaction price $Y_{t0}$. The continuity to the left
and right of $Y_{t0}$ is the bids and offers respectively, represented
by ${BB_{1},BB_{2},BB_{3},BB_{4}...}$ and ${BO_{1},BO_{2},BO_{3},BO_{4},BO_{5}...}$.
We assume that $Y_{t0}$ has exhausted the demand and supply at that
price. At $t_{0}$ on both sides of $Y_{t0}$ on the price axis, there
will exist a visible order book and a hidden latent order book. Such
order books are generally known to take a $V$ or $U$ shape. The
bid side demand is at an angle $\theta_{1}$ and the ask side offer
is at an angle $\theta_{2}$. This is a visual representation of the
V or U shape and a easy way to approximate the difference between
demand and supply if it exists. The difference between $\theta_{1}$
and $\theta_{2}$ generates a bias in the system, leading to the drift,
that we illustrate below. 

Intuitively, market orders want to take the market particles and hence
transactions towards $+\infty$ on Ask side and $-\infty$ on bid
side. The drift opposes this movement. Let us assume the introduction
of $M$, starts with $BO_{1}+BO_{1h}$, $BO_{2}$ that is visible
and $BO_{3h}$ which is hidden. The subscript $h$ is used for the
latent / hidden orders. Participant positions are based relative a
dynamic value such as the last transaction price (one may alternatively
reference either a fundamental value of the asset or an index etc).
The introduction of $M$ makes $\theta_{2}>\theta_{1}$, the magnitudes
of which is not known. The existence of large demand is revealed only
on the basis of visible order book and transactions. At $t=t_{1}$,
the transaction price moves to $Y_{t1}BO_{1}$. The block $M$ now
behaves like a sink or wall. It does not allow transactions to move
towards $BO_{2}$, $BO_{3}$. After $Y_{t1}BO_{1}$, a series of trades
can take place at $BO_{1}$ which is the new transaction price - $Y_{t2}BO_{1},Y_{t3}BO_{1}...Y_{t5}BO_{1}$.
Thus, successive transactions $Y_{t2},Y_{t3}...Y_{t5}$ are recorded
at the same price $BO_{1}$. The search for trades is still on in
both directions. Once market particles exhaust the liquidity at $BO_{1}$,
the trade again reaches $Y_{t0}$ which is the new best bid. New transaction
price is $Y_{t6}Y_{t0}$. If no liquidity is available here, the levy
search can move to either $BO_{1}$ or $BB_{1}$. Since, the former
keeps locking up the noise traders in limit orders, more likely the
trade moves to $BB_{1}$ within a couple of cycles. Hidden liquidity
can reappear at $Y_{t0}$ only if there is a readjustment of positions.
By this time, the market particles system has made sense of the gradient
and the drift can get activated. This simply involves correcting the
bias $\theta_{2}>\theta_{1}$.

\begin{figure}
\includegraphics[width=0.8\columnwidth]{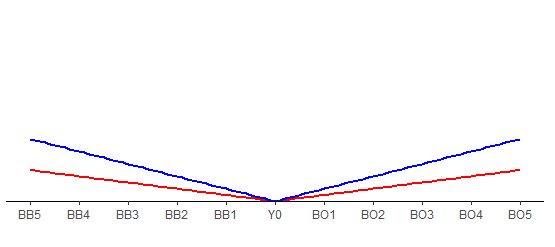}
\caption{Visual representation of a balanced total order book, visible orders shown in red and the cumulative
visible and hidden orders in blue. Hidden orders comprise the orders
that may be placed to demand or supply liquidity. The hidden limit
orders are placed at a later time and the hidden market orders are
placed at the particular transaction prices. The y-axis is the quantity.
$Y0$ is the last transaction price. The
bid side demand is at an angle $\theta_{1}$ and the ask side offer
is at an angle $\theta_{2}$. Here, $\theta_{1}=\theta_{2}$ . Fig 4 to Fig 8 is a sequential illustration 
showing biases due to new supply demand addition and resulting transactions 
and adjustments due to quote revision.}

\includegraphics[width=0.8\columnwidth]{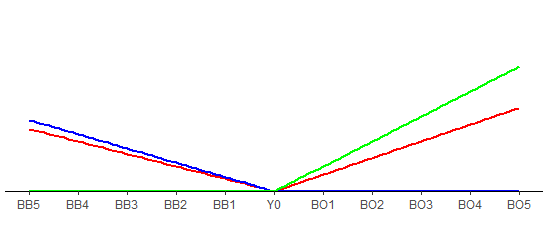}\caption{A bias develops around the last transaction price given in fig 4. Visible orders
shown in red and cumulative with hidden demand in blue and cumulative
with hidden supply in green. The y-axis is the quantity. $Y0$ is
the last transaction price.A large demand $M$ is added at BO1 (visible+hidden), BO2 (visible+hidden), B03 (hidden). The
bid side demand is at an angle $\theta_{1}$ and the ask side offer
is at an angle $\theta_{2}$. Here, $\theta_{2}>\theta_{1}$ a demand supply mismatch that leads to bias in the system due to hidden orders. (Fig 4 to Fig 8 is a sequential illustration showing biases due to new supply demand addition and resulting transactions and adjustments due to quote revision)}
\end{figure}

\begin{figure}
\includegraphics[width=0.8\columnwidth]{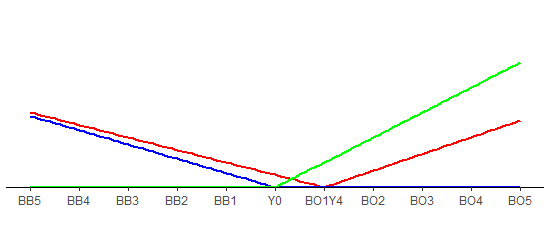}
\caption{After a few transactions from fig 5, the visible order book moves to new last transaction price $BO1Y4$ as reference
by adjustment of limit orders ($Y4$ is the fourth tansaction price). Visible orders shown in red and latent
demand in blue and latent supply in green. The latent order book is referenced to a fundamental price or the equilibrium price from the initial auction. $Y0$ is the first transaction price in the illustration. The y-axis is the quantity. (Fig 4 to Fig 8 is a sequential illustration showing biases due to new supply demand addition and resulting transactions and adjustments due to quote revision)}

\includegraphics[width=0.8\columnwidth]{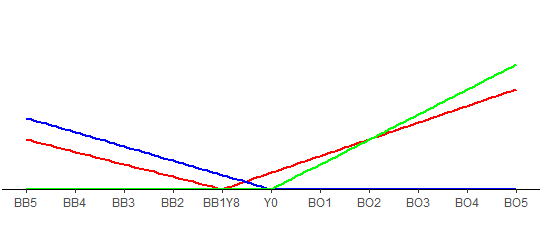}

\caption{A few transactions afterfig 6,  the visible order book moves to new transaction price $BB1Y8$ as reference
by adjustment of limit orders. The movement is from BO1 to BB1. Visible orders shown in red and hidden
demand in blue and hidden supply in green. Note the bias adjusting
in visible orders but the bias in hidden order remains with $\theta_{1}$
and $\theta_{2}$, the relative angles with the axis remaining the
same Fig 4 to Fig 8 is a sequential illustration showing biases due to new supply demand addition and resulting transactions and adjustments due to quote revision)}

\includegraphics[width=0.8\columnwidth]{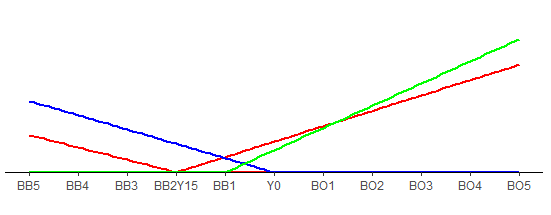}

\caption{The transaction price (Y15) moves further to BB2Y15 by the levy search.
Visible orders shown in red and hidden demand in blue and hidden supply
in green. The bias due to supply corrects as it dissociates from demand through quote revision
and moves the reference point to BB1 reducing $\theta_{2}$. (Fig 4 to Fig 8 is a sequential illustration showing biases due to new supply demand addition and resulting transactions and adjustments due to quote revision)}
 
\end{figure}

The correction of bias is the action based on readjustment of position
on either side. The reference point $Y_{t0}$ needs to move to $BO_{1},BO_{2}$
or $BB_{1},BB_{2}$. This is the drift in action and usually takes
place when participants revise their position on the basis of new
information or are not satisfied with the slow rate of trades. Once
transaction pace picks up on either side, there could be a new source
of drift from released market particles. These particles can get involved
in the carry trade for a small margin. In the absence of new information
and no adjustment on bid side, the transaction price gradually moves
up from $BB_{1}$, $BB_{2}$. The best offers start to move from $BO_{1}$ to
$Y_{t0}$, $BB_{1}$. In this situation, the demand $M$ can face
the lack of liquidity forcing a readjustment as illustrated in fig
6-8.

When $M>>G_{i}\omega(\chi)$ we can ignore the second term in the
equation. The expression for density reduces to:

\begin{equation}
g(x,t)=e^{v^{*}t}\frac{M}{\sqrt{4\pi DT}}exp[\frac{-1}{4DT}e^{-v^{*}t}(x-\int_{0}^{t}(\lambda_{2}+\lambda_{1}\phi_{2})ds)^{2}]
\end{equation}
In (28) , what happens if market orders are time dependent
and not simply as a response to one very large order. We assume this
large input of market orders is large enough to assume the initial
density to be very low So, $Q(x,t)=G_{0}f(0,t)\delta(x)$. Here the
function $f(0,t)$ shows the continuous availability of market orders
at a price. This could be in response to liquidity demand in the order
book in limit orders or the perceived hidden limit orders. This liquidity
is the source of large liquidity. Further from the equation below
for gradient observed by liquidity providers (refer Appendix D.9):
\[
f(x,t)=\frac{v(t)-v_{1}(t)}{\lambda_{1}(t)}x^{2}+\phi_{2}(t)x+c
\]
We infer that $Q=\phi_{2}(t)$. In such situations $\lambda_{2}$
can be ignored and $v_{1}$, $v-v_{1}$ can be reduced to $v*$ to
further simplify the equation. 

\begin{equation}
 g(x,t)=
\end{equation}
\[
 \beta_{1}\int_{0}^{t}\frac{\phi_{2}}{\sqrt{4\pi D(T-\zeta)}}exp[\frac{-1}{4D(T-\zeta)}e^{-v^{*}t}(\frac{x}{\beta(t)}-\int_{0}^{t}(\frac{\lambda_{2}}{\beta(t)}+\frac{\lambda_{1}\phi_{2}}{\beta(t)})ds)^{2}]
\]

\subsubsection{High reaction rate}

When the reaction rate is high, we evaluate the dynamics under the
further assumption of no addition of market orders through $q(x,t)$.
We expect the density $g$ at the transaction price to go down. This
is represented as (30), which is the second term in (26).

\begin{equation}
 g(x,t)=
\end{equation}
\[
 \beta_{1}(t)\int_{-\infty}^{\infty}\frac{1}{\sqrt{4\pi DT}}G_{i}\omega(\chi)exp(-\frac{1}{4DT}(\frac{x}{\beta}-\int_{0}^{t}(\frac{\lambda_{2}}{\beta}+\frac{\lambda_{1}\phi_{2}}{\beta})ds-\chi)^{2})d\chi
\]

Such a situation could arise if the market orders hit a liquidity
pool in limit orders and a large number of noise traders that were
earlier locked up in limit orders here get released. The liquidity
pool causing high transaction rates and release of noise traders creates
a sustainable transaction rate. The contribution to $v$ is coming
only from the reaction rate. $\phi_{2}=0$ and (30) reduces to (31).
\begin{equation}
g(x,t)=\beta_{1}(t)\int_{-\infty}^{\infty}\frac{1}{\sqrt{4\pi DT}}\omega(\chi)exp(-\frac{1}{4DT}(\frac{x}{\beta}-\int_{0}^{t}(\frac{\lambda_{2}}{\beta})ds-\chi)^{2})d\chi
\end{equation}

\subsubsection{Drift has no time dependence and low reaction rate}

In case, the drift is not time dependent and only spatially dependent
and no self generated drift resulting from release of noise traders.
$\lambda_{2}=0$ and the drift in response to resource gradient reduces
to (32)

\begin{equation}
\lambda_{1}(t)\frac{\partial f(x,t)}{\partial x}=\lambda_{1}\frac{\partial f(x,t)}{\partial x}=\lambda_{1}(\frac{v-v_{1}}{\lambda_{1}}x)+\lambda_{1}\phi_{2}(t)
\end{equation}

\subsubsection{Continuous spatial inflow of market orders that is not time dependent}

This is a case of liquidity available at every price in hidden market
orders. In such a case we need to find the equivalent of (29) that
is spatially dependent and not time dependent,
\begin{equation}
 g(x,t)=\beta_{1}(t)\int_{0}^{t}\int_{-\infty}^{\infty}\frac{(v-v_{1})\chi}{\sqrt{4\pi D(T-\zeta)}}exp(\frac{-1}{4D(T-\zeta)}(\frac{x}{\beta}-\chi)^{2})d\chi d\tau
\end{equation}

If dispersion  $D$ too is assumed to be time independent $D=D_{0}$
and we assume $v,v-v_{1}=v*$ because this is a small value, (33)
reduces to (34). ($D$ could take other forms too).

\begin{equation}
g(x,t)=e^{v^{*}t}\int_{0}^{t}\int_{-\infty}^{\infty}\frac{v^{*}\chi}{\sqrt{4\pi D_{0}(T-\zeta)}}exp(\frac{-1}{4D_{0}(T-\zeta)}(e^{v^{*}t}x-\chi)^{2})d\chi d\tau
\end{equation}

\subsubsection{At the limit of continuous auctions}

What happens as $t\rightarrow0$ immediately after
a transaction / auction has ended. Once a transaction price is
established new market orders come in at that price. These market
orders could be buy or sell orders or both. Other than such deposition
we assume cancellation and transaction do not occur at this instance.
Since, in this scenario $v=v_{1}=constant$, $\beta=1$ and $\beta_{1}=1$.
However, $T$ and ($T-\zeta$) does not exist and we cannot determine
the density of market order particles. We did not evaluate this limit
for the levy search. However, in our model of levy search process
we know that market orders start coming in at the last transaction
price with a velocity drawn from a distribution and that the process
exists. This will provide the result to the time derivative at that
point. It may be meaningful to find what happens in the inter-trade
interval between two trades. The average of this interval is precisely
the average trade duration. Let two consecutive trade transactions
take place at time $t_{1}$ and $t_{2}$ , $|t_{2}-t_{1}|=\tau$ .
What happens in this regime will depend on the dominant process occurring
in the market: diffusion / levy search, drift due to high transaction
rate, drift due to perceived resource gradient. We keep mixed processes
out of the scope of the current analysis.

In the diffusive regime, the contributions from other processes is
neglected. Further, further we assume $Q=0$ as presence of large
market orders will invoke other processes, $\beta=\beta_{1}=e^{v^{*}\tau}$,
$T$ reduces to $(2v^{*}e^{2v^{*}\tau})^{-1}$. In (35) $B$ is
the bid-ask spread. (35) reduces to (36) on simplifying. (36) gives
the relationship between the initial density of particles, diffusion
coefficient, spread and interauction time with the density of particles.
To understand the intuition in (36) we further assume that the liquidity
demand at $Y_{t_{2}}$ comes from limit orders and the market particles
provide liquidity. To find the maximum capacity during the inter-auction
trade (we assume availability of targets is not a constraint), let
us assume the ideal condition that every particle that engages in
search reaches the target. And that every new addition even in the
period $\tau$ is only a deposition that can go in and provide liquidity.
So both $D$ and $v^{*}$ are equal to $1$. Fig. 9 and Fig. 10 gives
multiplying factor to the intial density as a function of the spread
and interauction times.

\begin{equation}
g(Y_{t_{2}})=e^{v^{*}\tau}\frac{\omega(Y_{t_{2}})}{\sqrt{4\pi D(2v^{*}e^{2v^{*}\tau})^{-1}}}exp(\frac{-1}{4D(2v^{*}e^{2v^{*}\tau})^{-1}}(\frac{B}{e^{v^{*}\tau}})^{2})
\end{equation}

\begin{equation}
g(Y_{t_{2}})=\frac{\omega(Y_{t_{2}})}{\sqrt{4\pi D(2v^{*})^{-1}}}exp(2v^{*}(\tau-\frac{B^{2}}{4D}))
\end{equation}

Fig. 9 confirms that in an ideal best search scenario the density
exists. As the spread increases, the density is reduced. At spreads
of 0.2 and below the density converges to the initial density of particles
in interauction times of 0.5 seconds. Fig. 10 magnifys the fig. 9
around the low interauction times prevalent in liquid equity markets.
For the spreads of 0.2 and below it shows convergence around the constant
($1/\sqrt{2\pi}$) in (36). This suggest the exponent of $e$ is $0$
that is the expression $(\tau-\frac{B^{2}}{4D})\approx0$. Thus we
expect in a fairly ideal scenario 40 percent of the initial density
of market particles is able to trade. This demonstrates how the presence
of arbitrageurs and other traders, who contribute to the market particles, can
help increase the depth of the market. Note that this is the range
we would obtain every time the efficiency of the search is equal to
the contribution from the addition.

\begin{figure}
\includegraphics[width=0.45\paperwidth]{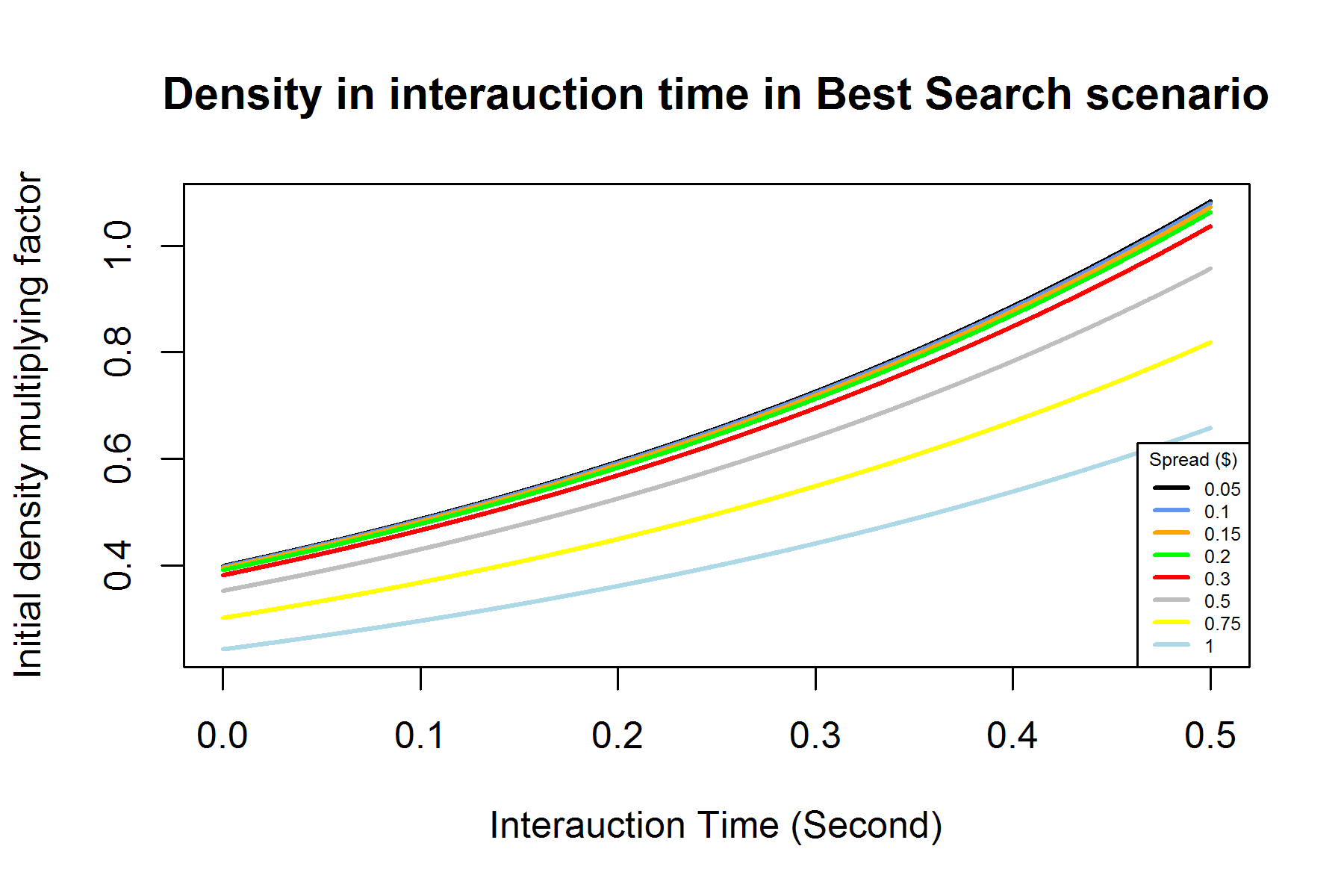}

\caption{The multiplying factor to the initial density as a function of the
interauction time ($\tau$) and spread ($B$). The multiplying factor gives the proportion of initial density of particles that are traded.}

\includegraphics[width=0.45\paperwidth]{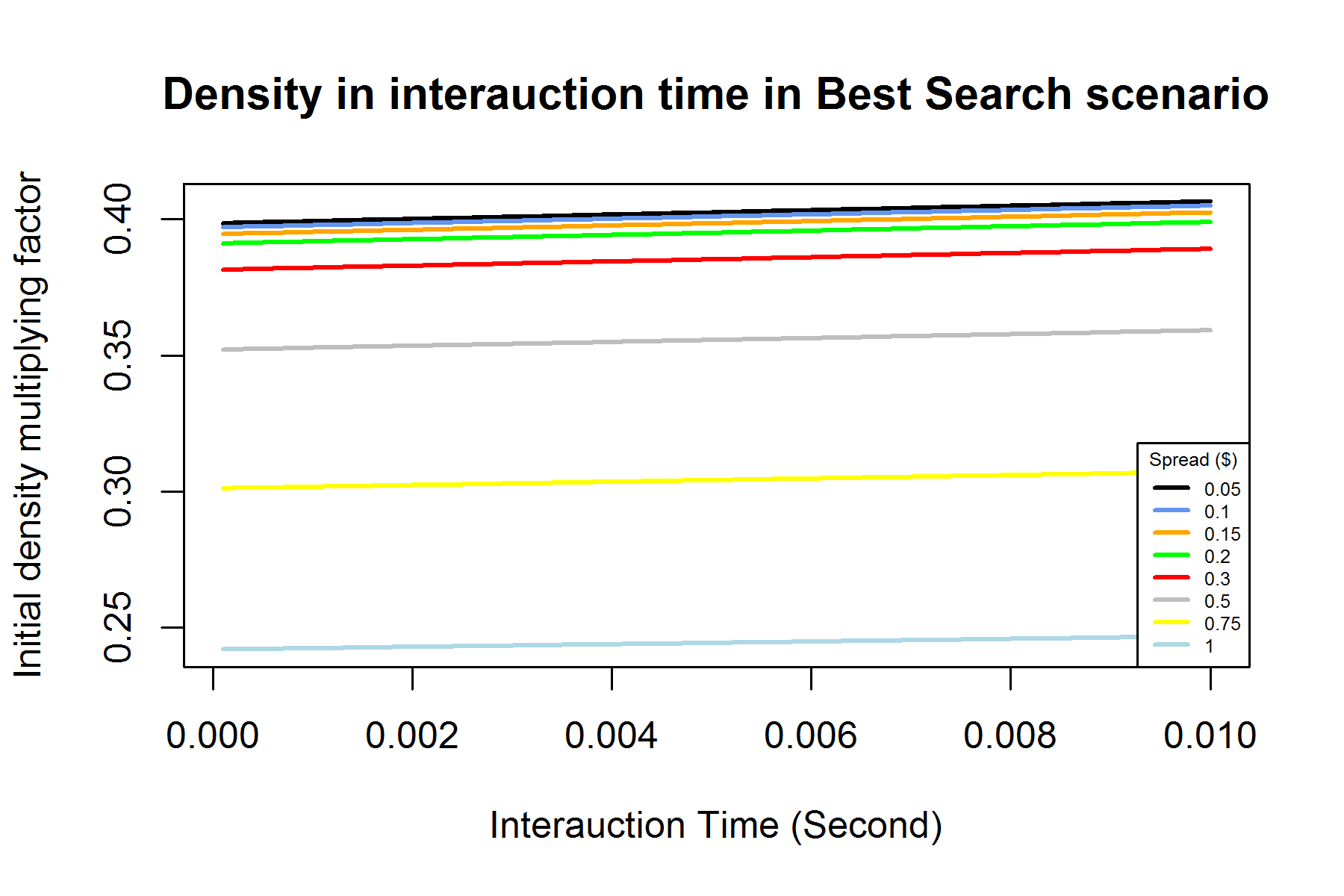}

\caption{The multiplying factor to the initial density as a function of the
interauction time ($\tau$) and spread ($B$). This figure magnifys
the fig 6(a) around the low interauction times prevalent in liquid
equity markets.}
\end{figure}

The relationship between the spread and diffusion coefficient can
be understood from the need for the following to hold: $B<2\sqrt{D}$
so that the exponent is positive. The diffusion coefficient could
be calculated in the standard way as MSD upon average flight time.

In the regime of drift due to high transaction rates, we ignore drift
due to any resource gradient. The scenario is that a large number
of market particles are already released due to high transaction rates
before we arrive at $t_{1}$. Since this component is significantly
higher than any new orders, $v_{1}>>v$, hence, $\beta=e^{-v^{*\tau}}$and
$\beta_{1}=e^{v^{*\tau}}$and $T=\frac{e^{2v^{*\tau}}}{2v^{*}}$.
In place of (36) we arrive at (37), where we can see that if the spread
is high the contribution from this drift will be small towards making  market particles available for transaction.

\begin{equation}
g(Y_{t_{2}})=\frac{\omega(Y_{t_{2}})}{\sqrt{4\pi D(\frac{1}{2v^{*}})}}exp(-\frac{2v^{*}}{4D}(B-\lambda_{2}^{*}\tau)^{2})
\end{equation}

When under the process of drift due to resource gradient, $v>>v_{1}$
and we neglect $v_{1}$. The equivalent expression is given in (38).
Again a high spread can reduce the contribution of this process in
the limit under consideration.

\begin{equation}
g(Y_{t_{2}})=\frac{\omega(Y_{t_{2}})}{\sqrt{4\pi D(\frac{1}{2v^{*}e^{2v^{*}\tau}})}}exp(-\frac{1}{4D(\frac{1}{2v^{*}})}(B-\lambda_{1}^{*}\phi_{2}^{*}\tau)^{2})
\end{equation}

\section{NUMERICAL SIMULATION}

We performed a numerical simulation of the model to improve our understanding
of the results. The numerical simulation is different than the regular
reaction-diffusion simulation because while orders may be present
in the total order book, they come into the limit order book at different
times. Researchers have in the past modelled the arrival of orders
into the limit order book using a poisson or a hawkes process (\citet{gould_limit_2013}).
Further, limit orders may get cancelled and leave the order book or
may get cancelled and get modified, the latter is usually treated
as a new order. 

\subsection{Pseudo-experiment set up}
We begin with a market where the number
of traders in the market is fixed at 1000. A trader places only a single order. An order is equivalent of a particle discussed in section 2. 

\subsubsection{Simulating Bias}
There are in all two sets of simulation that differ only in one aspect: in the first set, the quantity
of shares in the trader\textquoteright s order is one while in the
second set, it may vary between one to five. We use this difference in ordered quantity of shares to simulate the bias. The minimum order quantity is one share.

\subsubsection{Simulating Resource Density: order book events}
The arrival of events (order arrival, order cancellation, transactions)
is modelled following a poisson process. While transactions are not
orders, this seems to be a practical alternative to represent the
transactions in the tick by tick data as and when they occur. Thus,
an event of order arrival / order cancellation is followed by a order
matching to determine the possibility of transaction. If there is
a transaction, it is posted with a timestamp of the subsequent event
arrival time, else the next order arrives at the timestamp. The event
arrival rate is one of the determinants of the resource density. 

\subsubsection{Traders}
The percentage of noise traders in the market is assigned randomly in
each experiment drawn from a uniform distribution. The quoting and cancellation behaviour is governed
by the velocity and flight time distributions. The simulation assumes
a cauchy velocity distribution (ref Eq. 6) and a flight time distribution
with power tails (ref Eq. 7). The simulation is initialised with a
starting price or known price for the asset and the traders assigned
a velocity and a flight time, both drawn from the distribution. Traders are
randomly assigned as a noise or strategic trader. This is subject
to total number of noise traders in the experiment. The sign of the
velocity distinguishes the buyers and sellers. 

Type of order choice -- Limit or market is randomly assigned. The product of velocity
and flight time provides the deviation of price quote decision by
the trader from the initialised starting price. This is relevant for
the strategic traders who can start by placing limit orders, as noise
traders always start with market orders. Strategic traders exit the
market after the transaction or the orders are cancelled at the end
of the flight time. Noise traders after the initial placement, alternate
between market orders and limit order placement and exist in the market
till they are able to buy the asset with available funds or till end
of flight time. The pricing rule for limit order placement by Noise
traders is with a spread of 0.1 from the last transaction price. 

\subsubsection{Event arrival rates and experiments}
 Event arrival rates vary for thickly
versus thinly traded stocks. For each stock event arrival rates can
vary within the trading day.
The set of event arrival rates (measured as events per second) included in the experiment
is \{1,6,10,100,1000,10000\}. The included set covers a wide spectrum
of event arrival rates and hence the resource density. It covers the
necessary range, but is not dynamic i.e does not vary through the
trading period. The included set for $\gamma$ (power of flight time distribution from (7)) is \{0.25,0.5,0.75,1,1.25,1.5,1.75,2,2.25,2.5,2.75\}.
$\gamma$>2 is the region of Brownian search, 1< $\gamma$<2 is the
region of superdiffusive or levy search, $\gamma$<1 is the region
of ballistic search. 30 experiments each are simulated for each pair
of event arrival rate and $\gamma$. This design results in a total of 3960 experiments (two sets of 1980 experiments each for bias and non-bias case)

\subsection{Results}
Fig. 11 gives the results of the numerical simulation (no-bias case) showing the
rate of trades (identified as the search efficiency) per event arrival
rate. Since the quantity quoted and traded is one unit, there are
no biases arising due to resource density in this set of simulation.The
simulation could be further extended by including a trading model
and responses of traders to a bias. Such a simulation could demonstrate
the search efficiency of examples discussed in section 2.2.1-2.2.6.

\begin{figure}

\includegraphics[width=0.45\paperwidth]{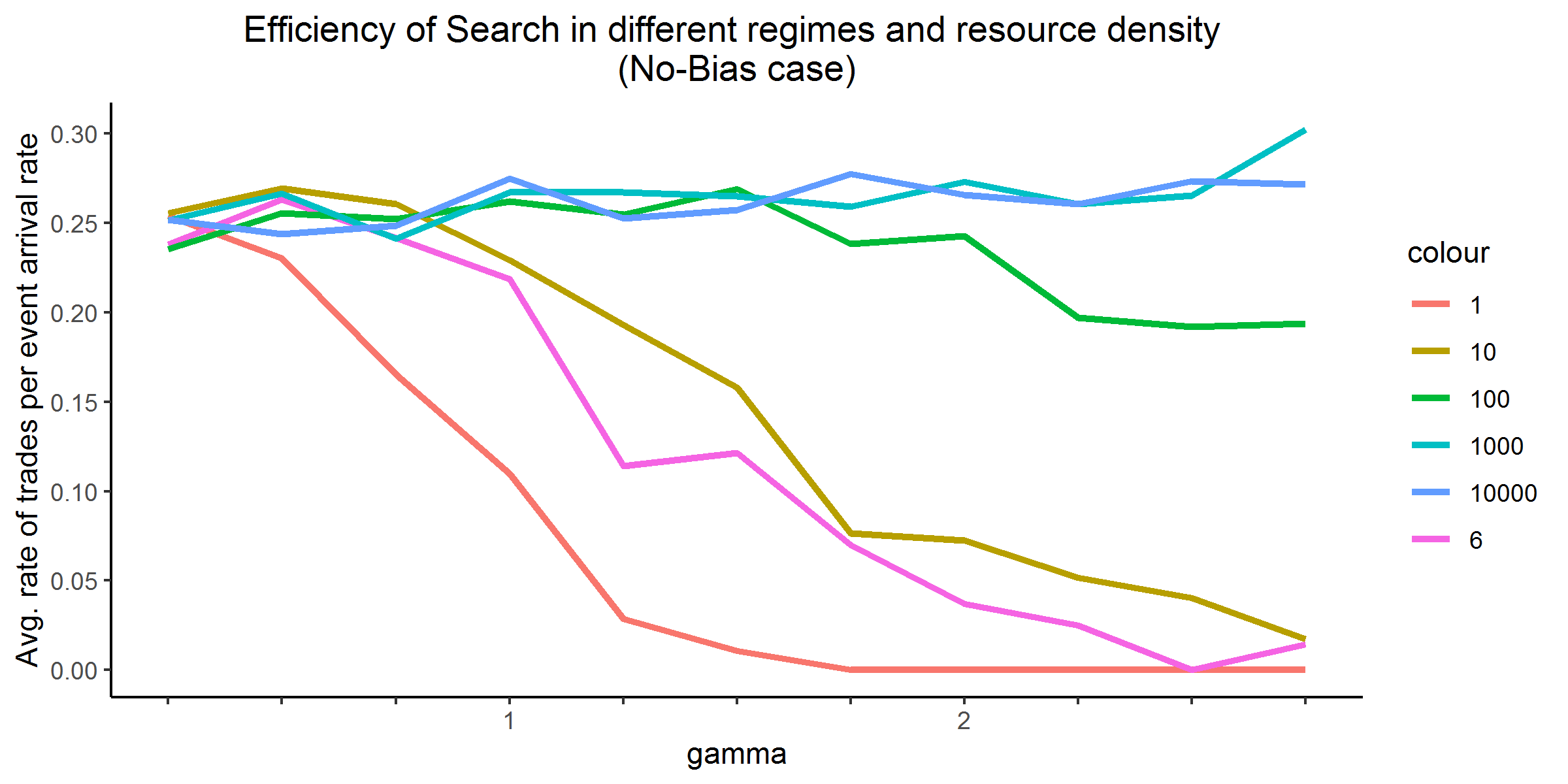}

\caption{The efficiency of search depends on the search regime and the resource
density. The former is set through the quoting and cancellation behaviour.
And the latter through the event arrival rates. In low resource density,
superdiffusive (1< $\gamma$<2 ) and ballistic search ($\gamma$<1)
perform better. When resource density increases and in absence of
any biases in the environment, brownian search is as efficient as
the others.}
\end{figure}

\begin{figure}
\includegraphics[width=0.45\paperwidth]{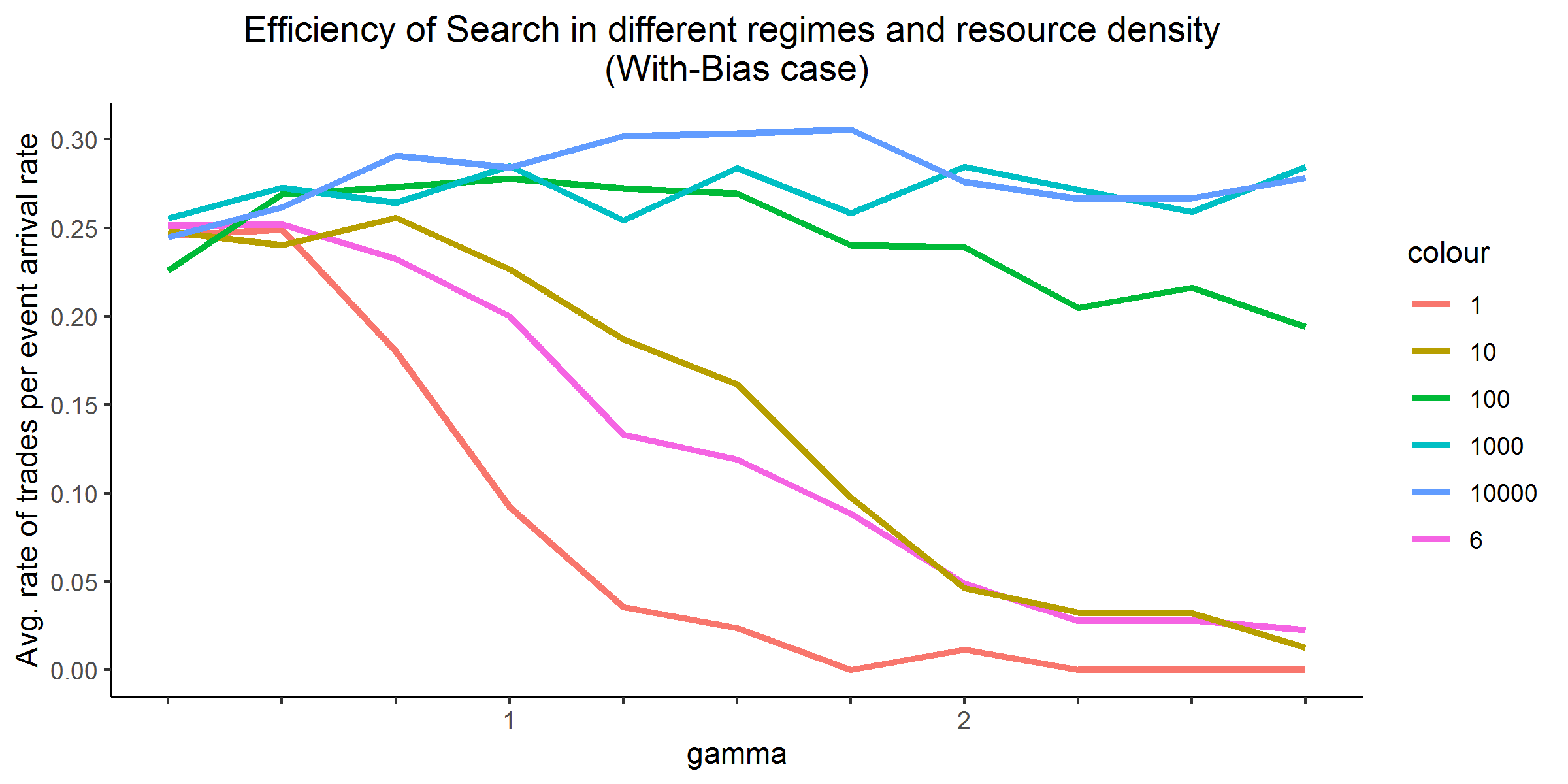}

\caption{The efficiency of search depends on the search regime and the resource
density. The former is set through the quoting and cancellation behaviour.
And the latter through the event arrival rates. Additionally, the
presence of Bias due to resource density increases efficiency of superdiffusive
search more than brownian search. In low resource density, superdiffusive
(1< $\gamma$<2 ) and ballistic search ($\gamma$<1) perform better. }

\end{figure}

Unlike ballistic search, search efficiency is dependent on the event
arrival rate (resource or target density) in both brownian and superdiffusive
search. This is in line with the findings in \citet{palyulina2014levy}.
When trading intensity is low (event arrival at 1 to 100 per second),
super-diffusive and ballistic search is more efficient than brownian
search with ballistic search being the most efficient. When the resource
density is high with event arrivals at 1000 events per second or higher,
both diffusive and superdiffusive search is equally efficient. In
the presence of bias (fig. 12) however, the superdiffusive search
begins to perform better even when the resource density is very high
(event arrival rates of 10000 per second). 

Fig. 13 gives a comparison of the average trading rates under the
three search regimes normalised to event arrival rates. Superdiffusive
search is more efficient than Brownian in the presence of bias. Brownian
search performs better in the absence of bias and in an environment
of high resource density. These are clearly related to active trading
periods of liquid stocks. We do not find the rate of trades being
impacted by the relative presence of noise traders (fig. 14). The
poisson arrival of events (traders and orders) into the market could be the possible reason.

Fig. 15 gives a sample of the trade prices obtained under different
event arrival rates, for the case when $\gamma=1.5$ and traders can
quote different quantities (bias in environment). The price time plots
describe the behaviour of prices. The saw-tooth property of transaction price equilibriation in double auctions markets (\citet{plot08}) can be observed (except with event arrivals at rate of 10000/sec).

\begin{figure}
\includegraphics[width=0.5\columnwidth]{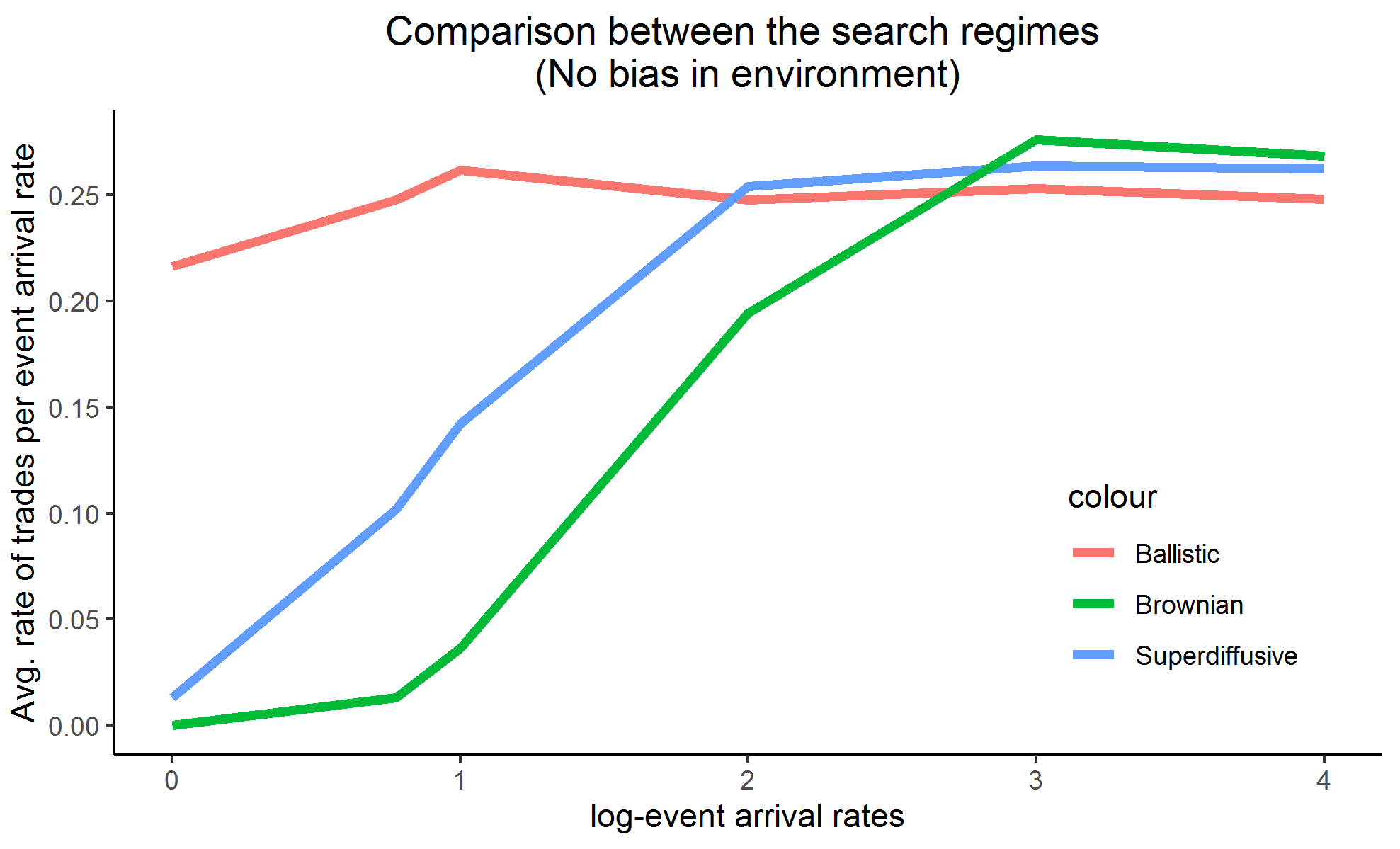}\includegraphics[width=0.5\columnwidth]{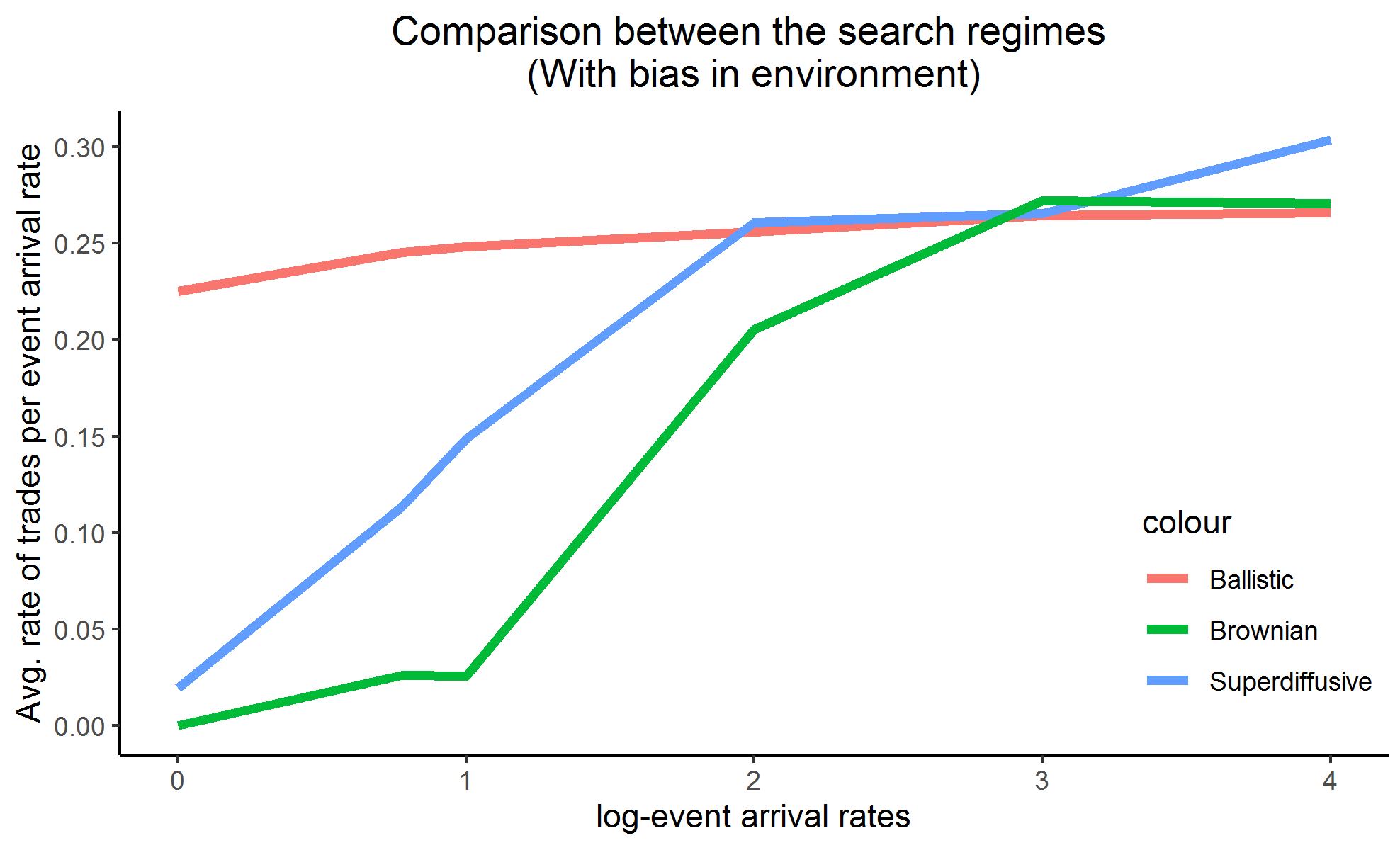}

\caption{Comparison of the search regimes in different resource density and
influence of bias. In the presence of bias, superdiffusive search
is more efficient than brownian. Ballistic search performs the best
in low resource density and is less influenced by resource density }
\end{figure}
\begin{figure}
\includegraphics[width=1.0\columnwidth]{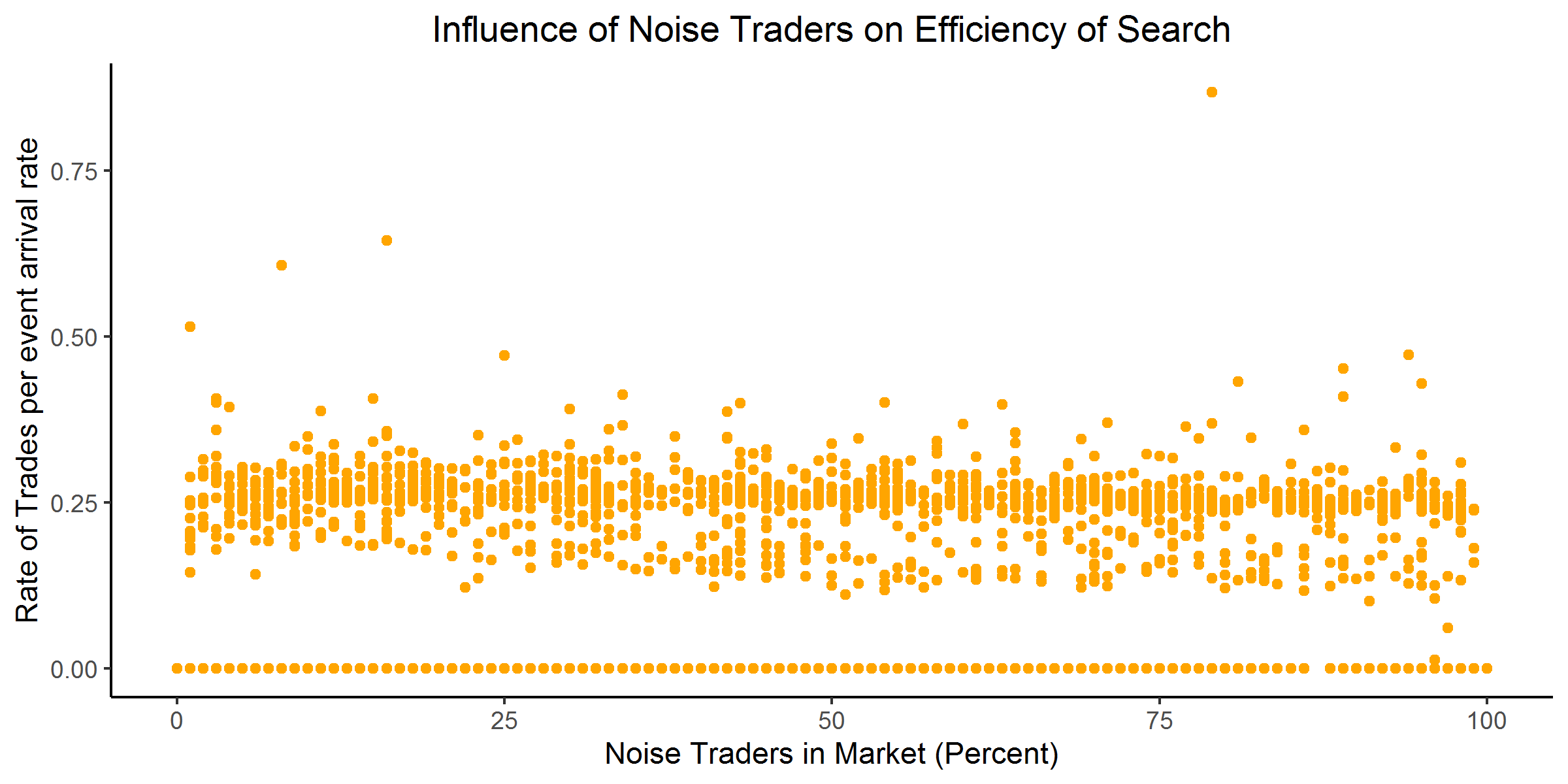}

\caption{The percentage of noise traders in each trial of the simulation is
randomly assigned. The number of noise traders does not affect the
efficiency of search. The possible reason could be the poisson arrival
of the traders. The above is the no-bias case that plots the rate
of trades per event arrival rate with the percentage of noise traders
in the market. The plot of the with-bias case is also similar.}
\end{figure}

\begin{figure}
\includegraphics[width=0.45\columnwidth]{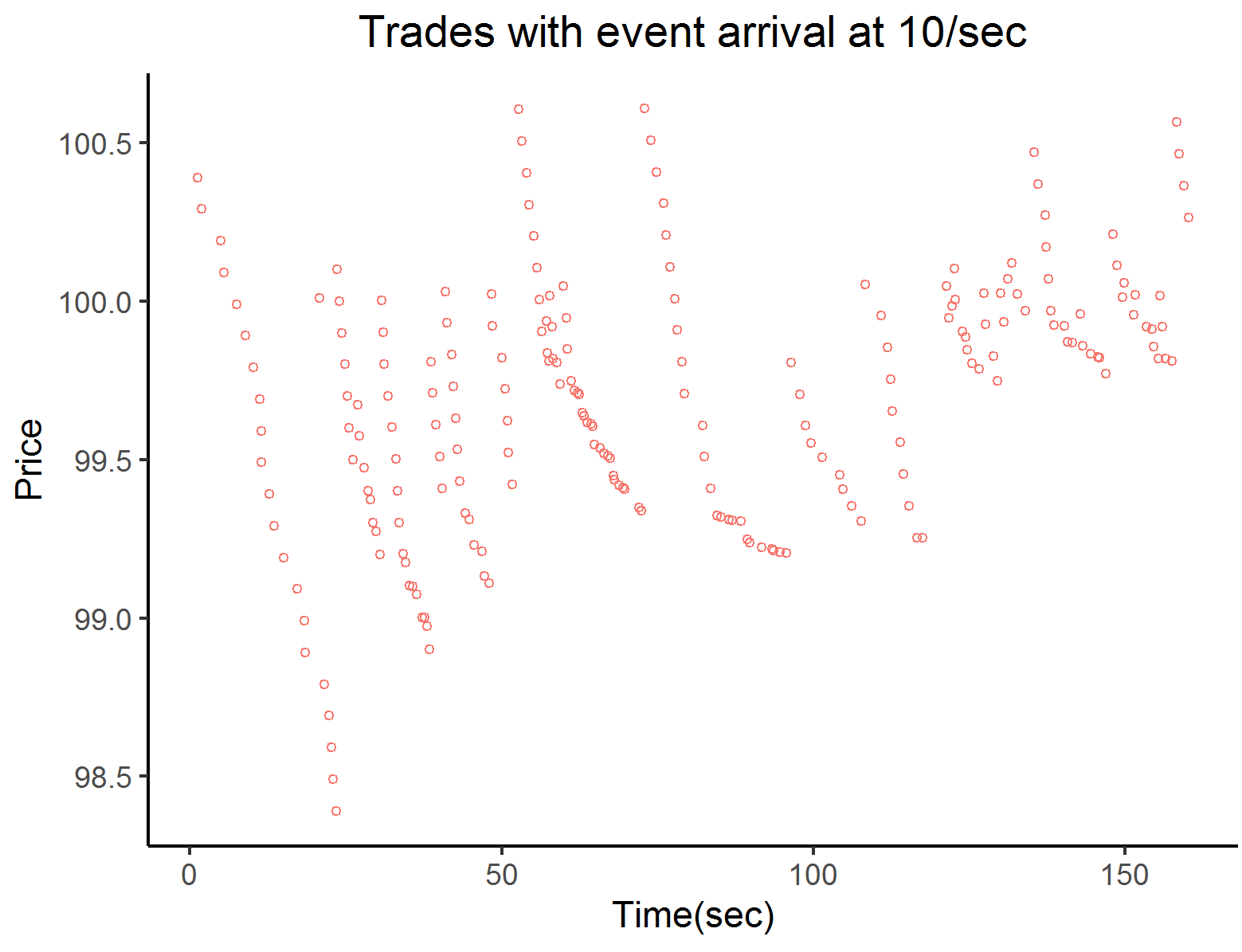}\includegraphics[width=0.45\columnwidth]{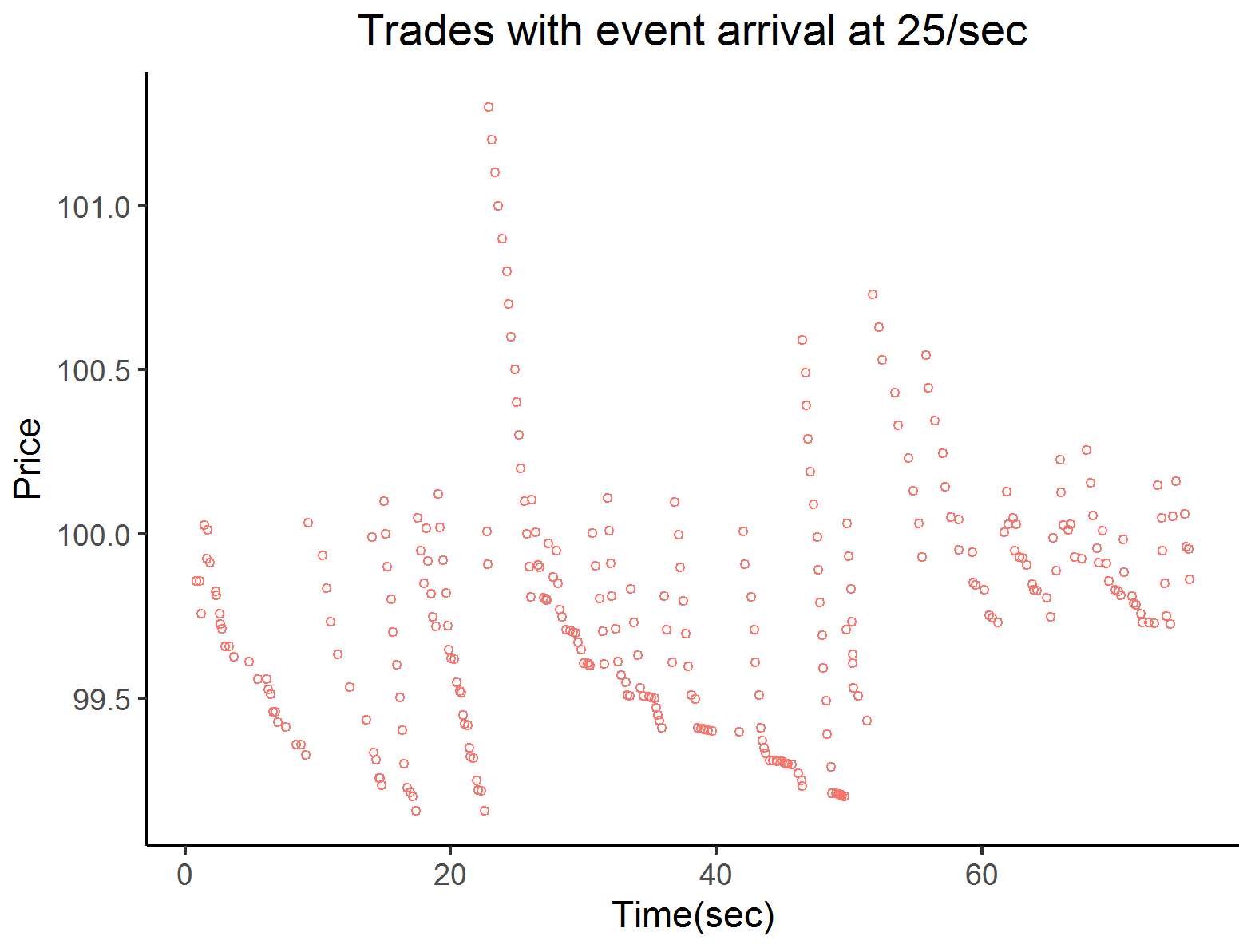}

\includegraphics[width=0.45\columnwidth]{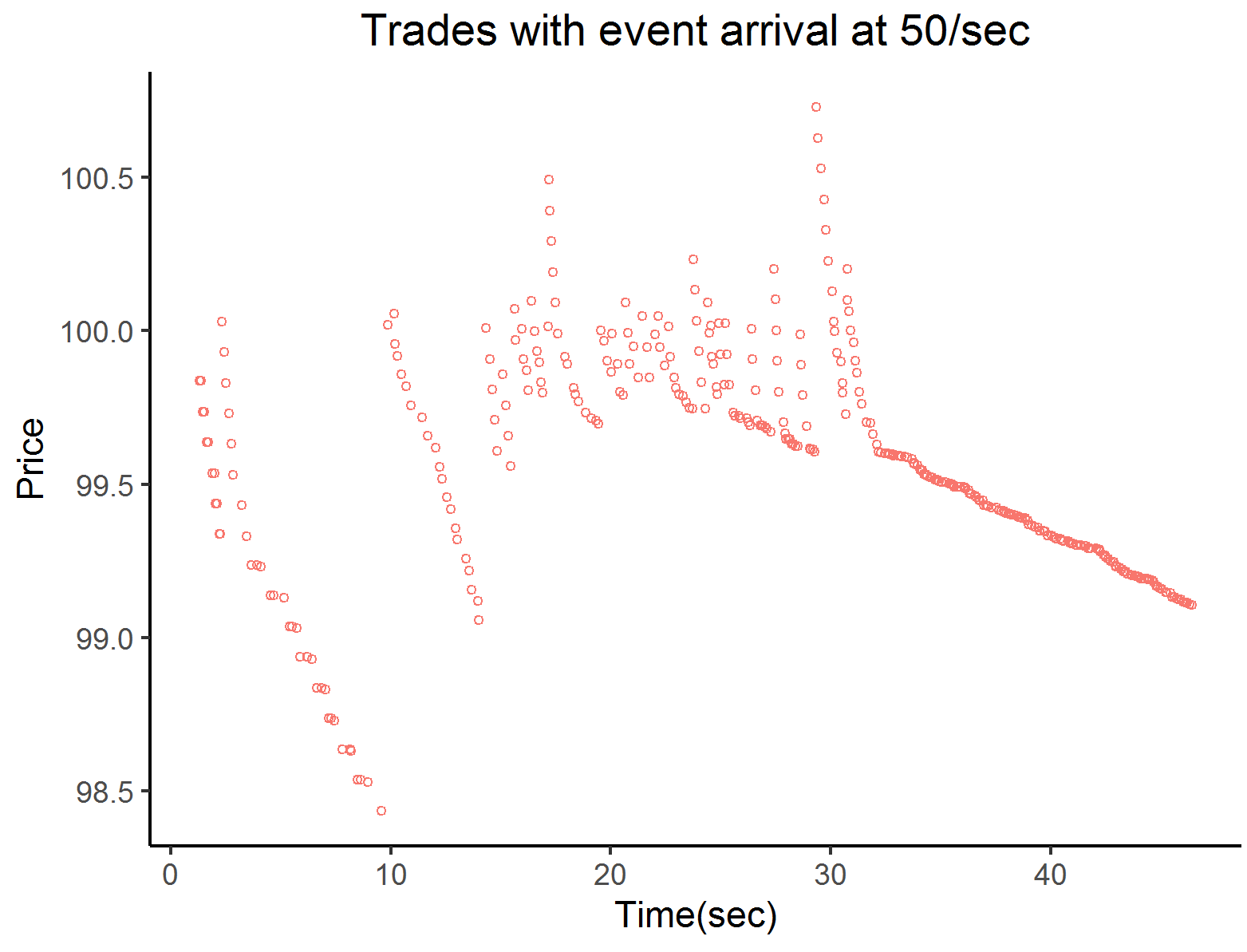}\includegraphics[width=0.45\columnwidth]{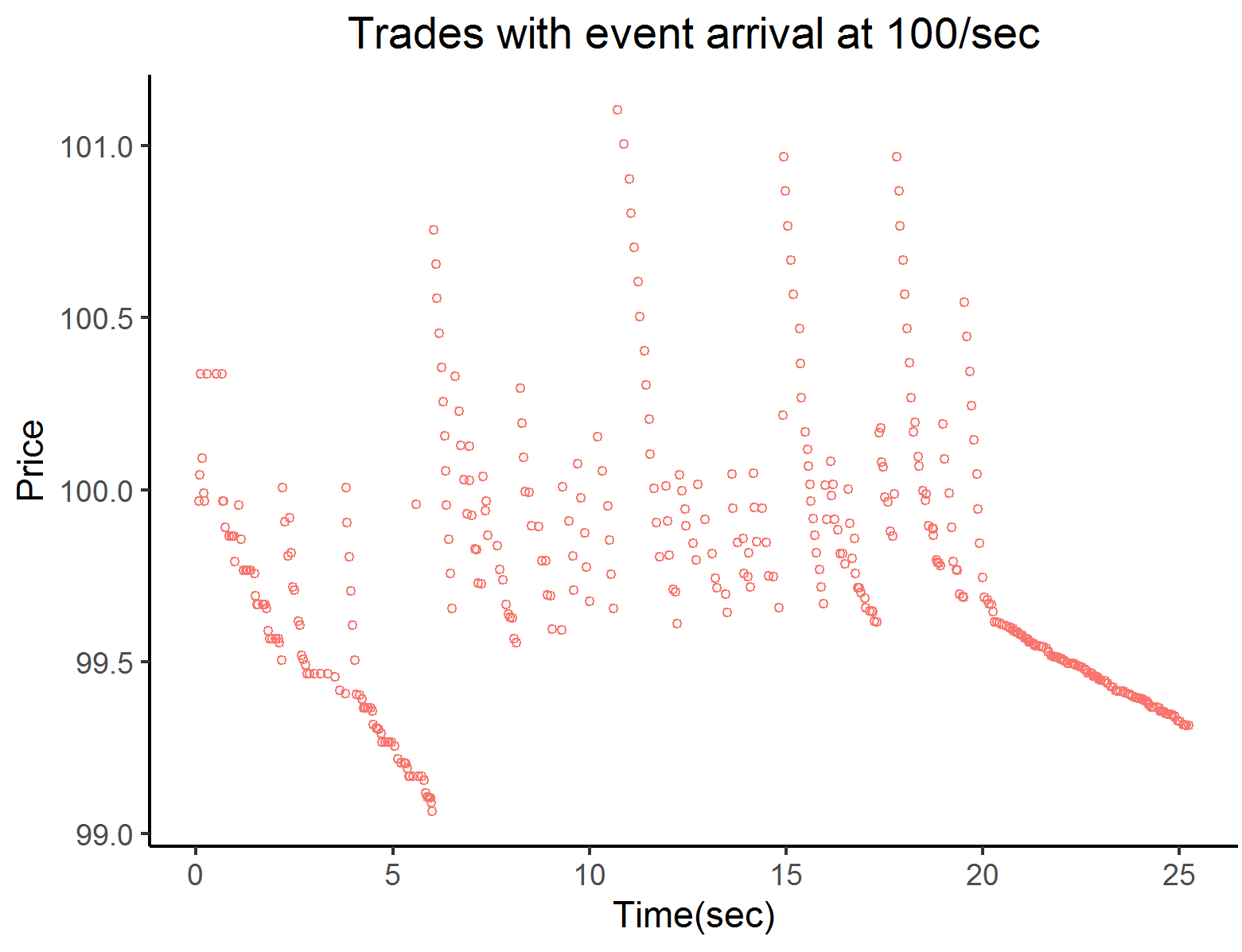}

\includegraphics[width=0.45\columnwidth]{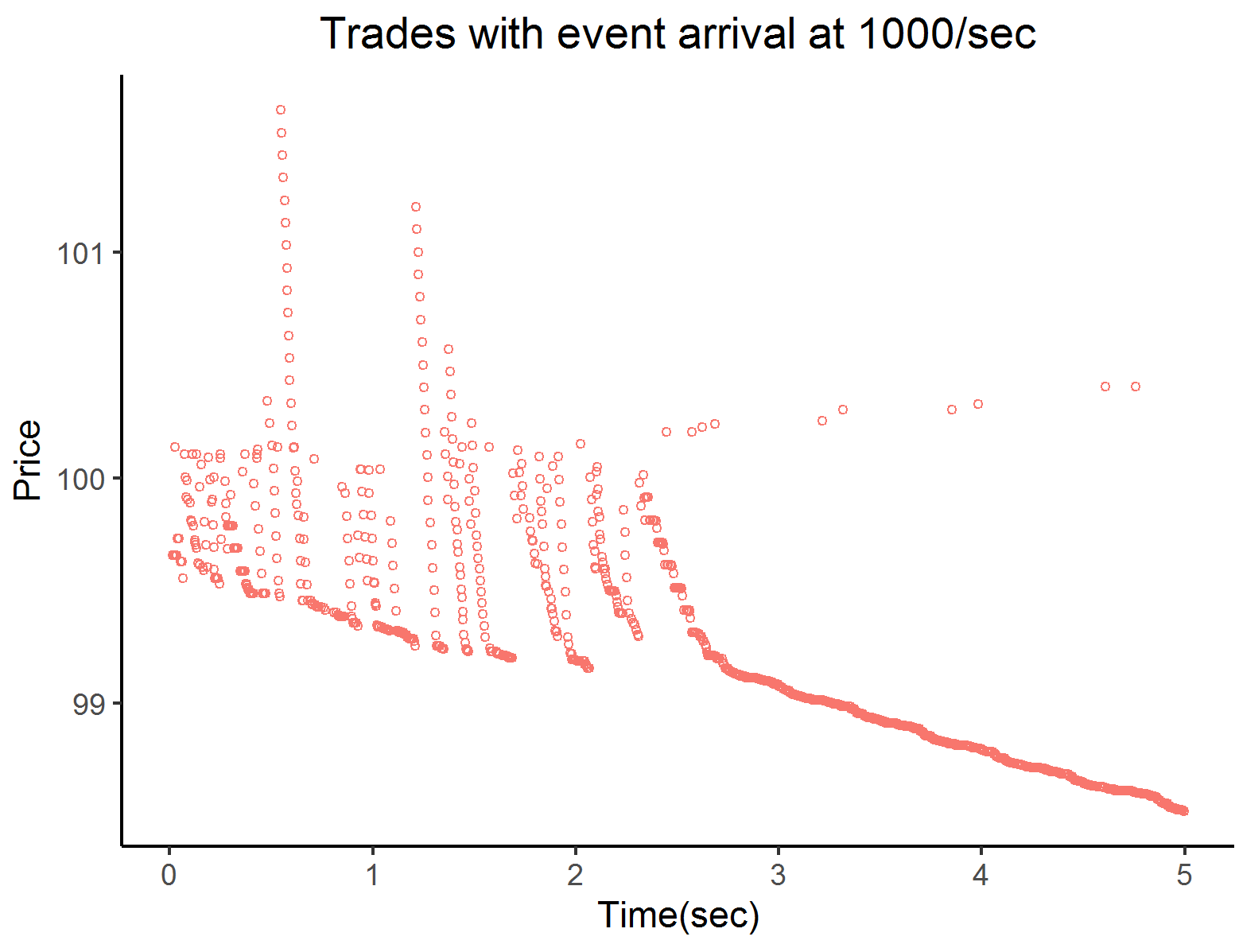}\includegraphics[width=0.45\columnwidth]{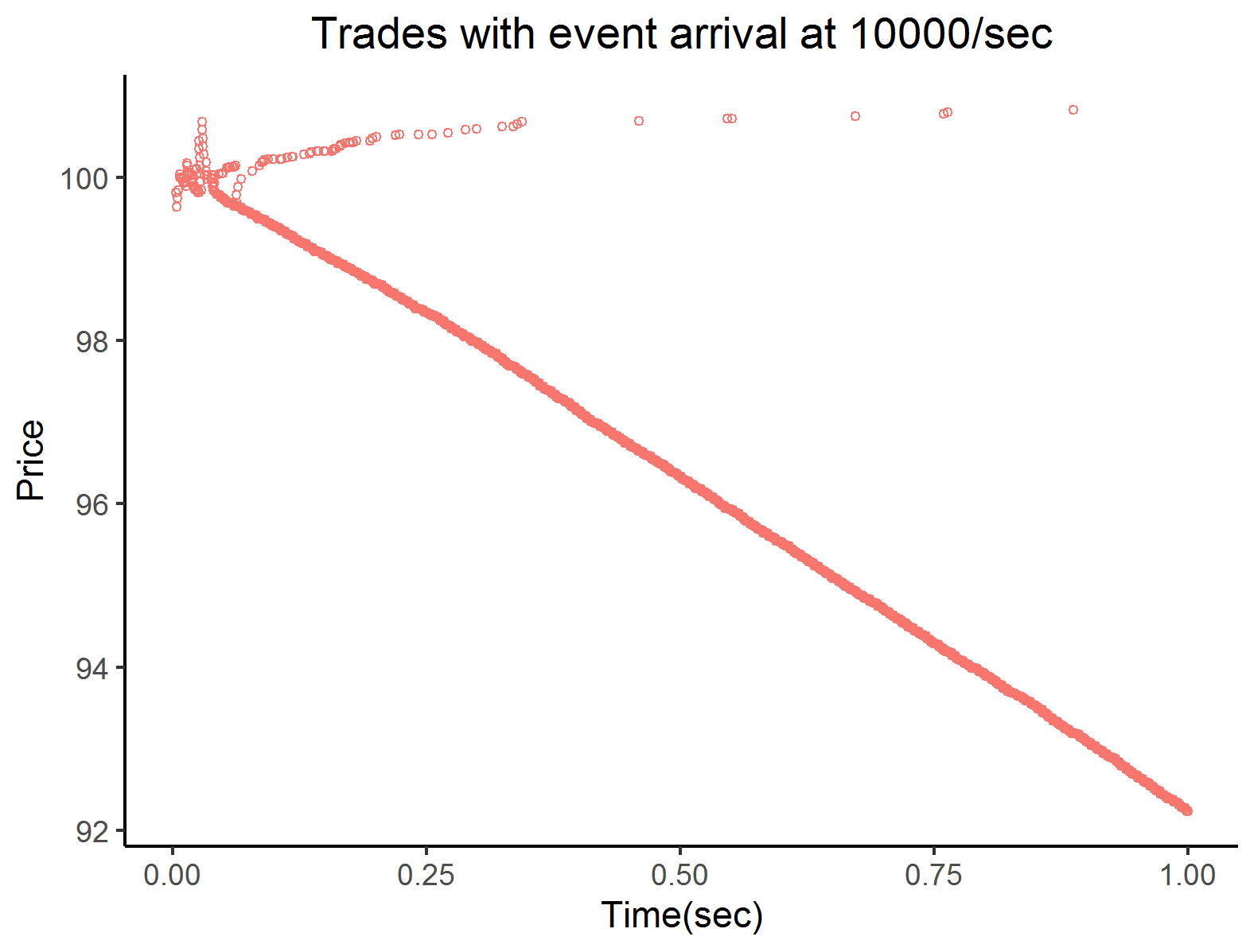}\caption{The behaviour of prices under the model. The plots of price and time
provides insight into the behaviour of prices. All samples are for
trade prices were obtained in the numerical simulation under an environment
of bias, i.e traders can quote different quantities. The parameter
$\gamma=1.50$, signifying superdiffusive search regime. The plots
represent various rates of event arrival - 10, 25, 50,100,1000, 10000
per second. The event arrivals includes order arrivals and a driver
of the resource density.}
\end{figure}

\section{DISCUSSION AND CONCLUSION}

We have introduced and analysed a model, to show how auctions in high
frequency markets without a designated market maker can be described
as a search by buyers for a seller and vice-versa. The model is based
on a zero intelligence approach. We consider a Total order book model
that includes a limit order book and the latent order book.

Intuitively, when the principle for matching trades is based on 'price-time
priority' in a continuous double auction mechanism, it seems to resemble a search. The need for the model arises
as existing theory based on supply and demand assumes the existence
or emergence of a walrasian equilibrium. The above assumption takes
the underlying model away from reality when considering intraday markets
that trade assets in high frequency through continuous auctions. In high resolution the assumption of an equilibrium of demand and supply leading to optimised quantity traded and emergence of price is fictional. This
is especially true in case of order driven markets without designated
market makers. The general belief that in financial markets given
that trading is continuous, prices can quickly adjust to clear the
market is valid but the process is not instantaneous. Information
flow is not seamless and frictions exist. Traders are not always present
in the market and alternate opportunities exist for the traders. Market
makers and dealers in quote driven markets were the key to push the
trading to equilibrium.

Our model is a diffusion-drift-reaction model and inspired by search
in biology and robotics. We analyse a number of asymptotic relationships
in the model. In the limit of continuous auctions we are unable to
determine the density of market particles or trading activity (similar
to \citet{donier_walras_2016}). We analyse the interauction times
in which the density exists and a relationship between the diffusion
coefficient, interauction time, initial density of particles and spread
exists. For the spread less than 0.2 USD and inter-auction time, 0.001
seconds or lower, less than 40 percent of the existing market particles
are able to trade. There exists a relationship between the spread
and the diffusion coefficient for an efficient market. So markets
that also depend on drift (resulting from high transaction rates involving
noise traders) apart from diffusion are adversely impacted when the
spread is high.

The numerical simulation run on the model brings out the search efficiency
of different search regimes (ballistic, superdiffusive and brownian)
in the simulated set up. The rate of trades that emerges from trade
duration is a natural candidate to measure the efficiency of search.
When biases exist due to resource density, the superdiffusive search
performs better than brownian search. In the absence of bias brownian
search is equally efficient in high resource density. In low resource
density ballistic search performs the best followed by superdiffusive
search. Performance of ballistic search is relatively not affected
by resource density or the presence of bias.

Future research can attempt to connect different elements of market
design, such as spread, tick size, presence of market maker etc, to
the search regimes to understand appropriate design basis the resource
density.

\section*{Acknowledgement}

We gratefully acknowledge valuable comments received, from an anonymous
referee on an earlier version of this paper and two anonymous referees on the current version, that has improved this
manuscript. 

\section*{}{}

\bibliographystyle{plainnat}
\bibliography{RefAuctionSearch}

\appendix{}
\section*{Appendix}

\section{The Search for trades in absence of bias}

Let $u(x,t)$ be the probability density function (pdf) of the distance
travelled by market order particles in search of trades. Thus, $x$
is the total length of the jumps and $t$ is the total time. The velocity
$v$ imparted by traders on particles can have positive and negative
values to include buy or sell orders (direction of motion). If $'N'$
number of trades takes place in a unit of time $dt$, the average
duration can be given by $\langle\mu\rangle=dt/N$. Each inter-trade
period is a search for the next transaction. A number of jumps get
recorded in this period. Equation (39) gives the total path traversed
in the search till time $t$. The initial distribution of the market
particles is given by $n_{0}(x)$.

\begin{equation}
u(x,t)=\int_{-\infty}^{\infty}dv\int_{0}^{t}u(x-v\tau,t-\tau)h(v)f(\tau)d\tau+n_{0}(x)\delta(t)
\end{equation}

The pdf of the traded particles, $q(x,t)$ is the number of trades
that take place in a time interval $(0,t)$. However, the trades being
the success of the search of market particles in a tatonnement, we
define the efficiency $\eta$ of the search process as the ratio of
the total path traversed in the search to the number of targets searched
or trades done. This gives us a link to represent $q(x,t)$ in terms
of $u(x,t)$ and $\eta$ as in equation (40). $\eta$ is again a
function of the coupling of velocity imparted and the flight times. 

\begin{equation}
q(x,t)=\int_{-\infty}^{\infty}dv\int_{0}^{t}\frac{u(x-v\tau,t-\tau)h(v)f(\tau)}{\eta(v\tau)}d\tau
\end{equation}

Equations (40) and (41) fully describe the dynamics of the system
with a given initial density of particles and the two pdf for the
flight times and velocities. They establish the crucial link between
the tatonnement in the auction with the trades. Next we solve these
equations analytically. First, we determine the total path traversed
by market particles in the search process and then introduce the result
into the equation of the trade density. We apply the Fourier transform
with respect to the spatial coordinate in (39). Due to the shift
property of the Fourier transform, an additional exponential factor
$e^{-ikv}$ appears under the integral. Integration with respect to
$v$ gives the Fourier transform of $h(v)$ with a reciprocal velocity
$k\tau$. The Fourier transform of (40) is:

\begin{equation}
u_{k}(t)=\int_{0}^{t}u_{k}(t-\tau)h_{k\tau}f(\tau)d\tau+n_{0,k}\delta(t)
\end{equation}

where the indices $k$ and $k\tau$ denote the Fourier components.
Note that, 
\begin{equation}
\mathcal{F}\{h(v)=\int_{-\infty}^{\infty}e^{-ik\tau v}h(v)dv\}=h(k\tau)
\end{equation}
. Next, we apply the Laplace transform with respect to time and use
its convolution property to obtain the following:
\begin{equation}
u_{k,s}=u_{k,s}[h_{k\tau}f(\tau)]_{s}+n_{0,k}
\end{equation}

where the index $s$ corresponds to the Laplace component. So the
path traversed in search can be given in the Fourier-Laplace domain,
$k,s$, as:

\begin{equation}
u_{k,s}=\frac{n_{0,k}}{1-[h_{k\tau}f(\tau)]_{s}}
\end{equation}

We can introduce the result in (44) into the Fourier-Laplace expression
for equation (41) (similar to above) to get the analytic expression
for the density of trades in the Fourier-Laplace domain, obtained
through the search process.

\begin{equation}
q_{k}(t)=\int_{0}^{t}\frac{u_{k}(t-\tau)h_{k\tau}f(\tau)}{\eta_{k}(\tau)}d\tau
\end{equation}

\begin{equation}
q_{k,s}=\frac{u_{k,s}[h_{k\tau}f(\tau)]_{s}}{\eta_{k,s}}
\end{equation}

\begin{equation}
q_{k,s}=\frac{n_{0,k}}{\eta_{k,s}}\frac{[h_{k\tau}f(\tau)]_{s}}{(1-[h_{k\tau}f(\tau)]_{s})}
\end{equation}

To find a solution, the next step is to take the Laplace inverse of
equation (47). However, an analytic representation and direct inversion
of the equation (47) is not feasible. \citet{froemberg_asymptotic_2015}
recommend an asymptotic analysis for large space and time scales,
$x,t\rightarrow\infty$. We use the same approach.
Going to Fourier-laplace space using the tauberian theorem, this limit
corresponds to $(k,s)\rightarrow(0,0)$ such that
$k/s=constant$. This has to be performed numerically.

It is important to define the velocity and flight time distributions
before we move to obtain the inverse transformation. In our view this
needs to come from empirical analysis. Further, velocity distribution
cannot be obtained directly as it is a notional quantity and needs
to be interpreted from equation (39) that describes the relationship
with returns.

As discussed in \citet{zaburdaev_random_2008}, \citet{froemberg_asymptotic_2015}
and \citet{zaburdaev_levy_2015} when the velocity distribution is
Cauchy or lorentian, the density of the particles also is lorentian
independent of flight times and jump lengths. Such a lorentian velocity
profile appears in real physical phenomena such as two dimensional
turbulence and is also found in model distributions of kinetic theory,
statistics, plasma physics and starving amoeba cells. We know that
a cautchy process does not give rise to a continuous sample path for
the price and it differs from Brownian motion as there are large jumps
not infrequently. As given below we make arbitrary choice of a lorentian
velocity distribution $h(v)$ and intuitively a flight time distribution
$f(\tau)$ with power tails.
\begin{equation}
h(v)=\frac{1}{u_{0}\pi}\frac{1}{(1+(\frac{v^{2}}{u_{0}^{2}}))}
\end{equation}

\begin{equation}
f(\tau)=\frac{\gamma}{(1+\tau)^{1+\gamma}}
\end{equation}

In equation (48) $u_{0}$ is needed to constrain the velocities
so that the particles do not go beyond the ballistic cones, else it
will lead to instantaneous dispersion. The $\gamma$ in equation (49)
is varied to get different transport. $\gamma=1/2$ for normal diffusion.
In equation (47), given the asymptotic limit we want to evaluate,
we further set a constant efficiency, so that there is an expression
for the initial density of particles for the trade density. The propagator
for our model can then be noted as follows:

\begin{equation}
G(k,s)=\frac{\mathcal{L}[f(\tau)h(k\tau)]}{1-\mathcal{L}[f(\tau)h(k\tau)]}
\end{equation}

where $k\tau$ is the Fourier variable conjugate to $v$. The equation
(50) retains the form of the well known Montroll-Weiss equation
for the pdf of the uncoupled continuous time random walk (CTRW) to
find the particle $x$ at the time $t$, modified such that it applies
to random jumps in velocity. Equation (50) can be rewritten as (see
Appendix C for details):
\begin{equation}
G(k,s)=\frac{\intop_{-\infty}^{\infty}dvf(s+ikv\tau)h(v)]}{1-\intop_{-\infty}^{\infty}dvf(s+ikv\tau)h(v)]}
\end{equation}

For the flight time distribution we have chosen and in the long time
limit the expansion in the Laplace space is given by,
\begin{equation}
f(\tau)\approxeq1-\tau^{\gamma}\Gamma(1-\gamma)s^{\gamma}
\end{equation}

Using (52) the asymptotic version of (51) is,

\begin{equation}
G(k,s)=\frac{1}{s}\frac{\intop_{-\infty}^{\infty}(1+ikv/s)^{\gamma-1}h(v)dv}{\intop_{-\infty}^{\infty}(1+ikv/s)^{\gamma}h(v)dv}
\end{equation}

\section{Inversion of the fourier laplace expression for propagators with
ballistic scaling}

A method exists for the inversion of the fourier laplace expression
for propagators with ballistic scaling. A propagator of a random walk
model has ballistic scaling if it can be written in the form, $G(x,t)\approxeq\frac{1}{t}\phi(\frac{x}{t}),\,t\rightarrow\infty$,where
$\phi$ is the scaling function. In Fourier-laplace space this is,
$G(k,s)\approxeq\frac{1}{s}g(\frac{ik}{s})$. Comparing the above
two forms we can rewrite the scaling form of our equation as in equation
(54), where $\xi=\frac{ik}{s}$ . 
\begin{equation}
g(\xi)=\frac{\intop_{-\infty}^{\infty}(1+\xi v)^{\gamma-1}h(v)dv}{\intop_{-\infty}^{\infty}(1+\xi v)^{\gamma}h(v)dv}
\end{equation}

Further, this can be inverted according to the general equation (55)
(from \citet{froemberg_asymptotic_2015} based on \citet{godreche2001statistics})
to obtain equation (56). The scaling function in (55) is defined
as $\phi(y)=\langle\delta(y-Y)\rangle$. Here angular brackets denote
the averaging with respect to a random variable $X$ which has a pdf
$P(X)$, $\langle F(X)\rangle=\intop_{-\infty}^{\infty}F(X)P(X)dX$
and $Y$ is the time average of the particles velocity, i.e $x/t$.
(55) is obtained using Sokhotsky-Weirstrass theorem: $\frac{\lim}{\epsilon\rightarrow0}\frac{1}{x\pm i\epsilon}=\frac{1}{x}\mp i\pi\delta(x)$
and thus $\mp\frac{1}{\pi}Im\frac{\lim}{\epsilon\text{\ensuremath{\rightarrow}}0}\frac{1}{x\pm i\epsilon}=\delta(x)$.

\begin{equation}
\phi(y)=\frac{-1}{\pi}\frac{\lim}{\epsilon\rightarrow0}\Im[\frac{1}{y+i\epsilon}g(\frac{-1}{y+i\epsilon})]
\end{equation}

\begin{equation}
\phi(y)=\frac{-1}{\pi}\frac{\lim}{\epsilon\rightarrow0}\Im[\frac{\intop_{-\infty}^{\infty}(1+i\epsilon-v)^{\gamma-1}h(v)dv}{\intop_{-\infty}^{\infty}(1+i\epsilon-v)^{\gamma}h(v)dv}]
\end{equation}

While the velocity distribution could be anything from a two state,
or uniform distribution (\citet{froemberg_asymptotic_2015} discuss
a number of examples) we arbitrarily choose the special situation
induced by Cauchy distributed velocity. To use this we start with
the following propagator and velocity distribution . The propagator
reduces to (59) without prescribing to any particular form for $f(\tau)$.

\begin{equation}
G(k,s)=\frac{\{f(\tau)h(k\tau)\}}{1-\{f(\tau)h(k\tau)\}}
\end{equation}

\begin{equation}
h(k\tau)=exp(-u_{0}|k|\tau)
\end{equation}

\begin{equation}
G(k,s)=\frac{1}{s+u_{0}|k|}
\end{equation}

Now taking the inverse Laplace and Fourier transform we get a form
of Cauchy distribution. The scaling function is $\phi(y)=\frac{1}{\pi(1+y^{2})}$.

\begin{equation}
G(x,t)=\frac{u_{0}t}{\pi(u_{0}^{2}t^{2}+x^{2})}
\end{equation}

\section{Explanation for equation (9)}

Numerator,
\[
\mathcal{L}[f(\tau)h(k\tau)]_{(k,s)}
\]

\[
=\intop_{-\infty}^{\infty}d\tau\int_{-\infty}^{\infty}e^{-ik\tau v}e^{-s\tau}f(\tau)h(v)]
\]

\[
=\intop_{-\infty}^{\infty}d\tau\int_{-\infty}^{\infty}e^{-(s+ikv)\tau}e^{-s\tau}f(\tau)h(v)]
\]

\[
=\intop_{-\infty}^{\infty}d\tau\int_{-\infty}^{\infty}f(s+ikv)h(v)]
\]

Similarly the denominator can be arrived at.

\section{Search for trades in presence of bias (The Complete Model)}

We build a comprehensive model of stochastic evolution of market particles.
Our fundamental set up of the complete model is (61) which is expanded
to (62)
\begin{equation}
\frac{\partial g(x,t)}{\partial t}=
\end{equation}
\[
\frac{\partial}{\partial x}(D(t)\frac{\partial g(x,t)}{\partial x}-\lambda_{1}(t)\frac{\partial f(x,t)}{\partial x}g(x,t)-\lambda_{2}(t)g(x,t))+v(t)g(x,t)+q(x,t)
\]

\begin{equation}
\frac{\partial g(x,t)}{\partial t}=D(t)\frac{\partial^{2}g(x,t)}{\partial x^{2}}-\lambda_{1}(t)\frac{\partial f(x,t)}{\partial x}\frac{\partial g(x,t)}{\partial x}-\lambda_{1}(t)\frac{\partial^{2}f(x,t)}{\partial x^{2}}g(x,t)
\end{equation}
\[
-\lambda_{2}(t)\frac{\partial g(x,t)}{\partial x}+v(t)g(x,t)+q(x,t)
\]

(62) can be solved analytically. We draw upon the technique used
by \citet{sanskrityayn_analytical_2016}, who used the Greens function
method to solve their diffusion-advection equation in the context
of pollutant solutes in the atmosphere. To solve the equation, we
note that $q$ will remain untouched and we need to reduce the equation
to a known form so that we find an expression for $f(x,t)$. We do
a co-ordinate transformation from the domain $(x,t)$ to the domain
$(X(x,t),t')$. The domain X is essentially fixed time-snapshots of
the entire lattice. We want to transform (62) to the form in equation
(63).

\begin{equation}
\frac{\partial G(X,t')}{\partial t'}=D_{1}(t')\frac{\partial^{2}G(X,t')}{\partial X^{2}}-\lambda(t')\frac{\partial G(X,t')}{\partial X}+v_{1}(t')G(X,t'))+q_{1}(X,t')
\end{equation}

Using the domain transformation, we can write equation (62) as equation
(64), following which we equate the coefficients to obtain equation
(65,66 and 67):
\begin{equation}
\frac{\partial G(X,t')}{\partial t'}=D(t')(\frac{\partial X}{\partial x})^{2}\frac{\partial^{2}G(X,t')}{\partial X^{2}}
\end{equation}

\[
-[-D(t')\frac{\partial^{2}X}{\partial x^{2}}+\lambda_{1}(t')\frac{\partial f(x,t)}{\partial x}\frac{\partial X}{\partial x}+\lambda_{2}(t)\frac{\partial X}{\partial x}+\frac{\partial X}{\partial t}]\frac{\partial G(X,t')}{\partial X}
\]

\[
+[v(t')-\lambda_{1}(t')\frac{\partial f^{2}(x,t)}{\partial x^{2}}]G(X,t'))+q(X,t')
\]

\begin{equation}
D(t')(\frac{\partial X}{\partial x})^{2}=D_{1}(t)
\end{equation}

\begin{equation}
-[-D(t')\frac{\partial^{2}X}{\partial x^{2}}+\lambda_{1}(t')\frac{\partial f(x,t)}{\partial x}\frac{\partial X}{\partial x}+\lambda_{2}(t)\frac{\partial X}{\partial x}+\frac{\partial X}{\partial t}]=-\lambda
\end{equation}

\begin{equation}
v(t')-\lambda_{1}(t')\frac{\partial f^{2}(x,t)}{\partial x^{2}}=v_{1}
\end{equation}

From (65) we obtain an expression for $X$ in (68) and from (67)
an expression for $f$ in (69). We insert these two results into
(66) and equate similar coefficients to obtain (70) and (71).
For the sake of convenience, we use t instead of t' hence forth.

\begin{equation}
X=\sqrt{\frac{D_{1}}{D}}x+\phi_{1}(t)
\end{equation}

\begin{equation}
f(x,t)=\frac{v(t)-v_{1}(t)}{\lambda_{1}(t)}x^{2}+\phi_{2}(t)x+c
\end{equation}

\begin{equation}
-\lambda_{1}(t)(\frac{v(t)-v_{1}(t)}{\lambda_{1}(t)}x+\phi_{2}(t))\sqrt{\frac{D_{1}(t)}{D(t)}}-\lambda_{2}(t)\sqrt{\frac{D_{1}(t)}{D(t)}}-\frac{\partial}{\partial t}(\sqrt{\frac{D_{1}(t)}{D(t)}}x+\phi_{1})
\end{equation}
\[
=-\lambda(t)
\]

\begin{equation}
\frac{\partial}{\partial t}\phi_{1}(t)-\lambda_{2}(t)\sqrt{\frac{D_{1}(t)}{D(t)}}-\lambda_{1}(t)\phi_{2}(t)\sqrt{\frac{D_{1}(t)}{D(t)}}=-\lambda(t)
\end{equation}

\begin{equation}
-(v(t)-v_{1}(t))\sqrt{\frac{D_{1}(t)}{D(t)}}=\frac{\partial}{\partial t}\sqrt{\frac{D_{1}(t)}{D(t)}}
\end{equation}

From (72), we get (73) where $\beta$ is a dimensionless expression
defined in (74)
\begin{equation}
\frac{D_{1}(t)}{D(t)}=\frac{1}{\beta^{2}(t)}
\end{equation}

\begin{equation}
\beta=e^{\int_{0}^{t}(v(s)-v_{1}(s))ds}
\end{equation}

Using the expression for $\phi_{1}(t)$, obtained after reorganising
(71), we obtain the expression for X as below,
\begin{equation}
X=\frac{x}{\beta(t)}+\int_{0}^{t}(\lambda(t)-\frac{\lambda_{2}(t)}{\beta(t)}-\lambda_{2}(t)\frac{\phi_{2}(t)}{\beta(t)})
\end{equation}

Equipped with the expression we have obtained for $f(x,t)$ in (69)
and the transformations $\beta(t$) and $X$ in (75) we can reduce
our initial equation to the following form,
\begin{equation}
\frac{\partial G(X,t)}{\partial t}=\frac{D(t)}{\beta^{2}(t)}\frac{\partial^{2}G(X,t)}{\partial X^{2}}-\lambda(t)\frac{\partial G(X,t)}{\partial X}+v_{1}G(X,t)+q_{1}(X,t)
\end{equation}

The initial conditions for this equation are $G(X,0)=G_{i}\omega(X),$
with $-\infty<X<\infty$ and $t>0$ . Next, we try to remove the drift
term and the decay term. We use the following transformation equations
one after the other for this purpose. In (78) $\beta_{1}=e^{\int_{0}^{t}v_{1}(s)ds}$
is a dimensionless term and in (79) $\beta$ is as per (73). In
(79) T is a time variable. Equation (76) now reduces to equation
(80).

\begin{equation}
\eta=X-\lambda(t)t
\end{equation}

\begin{equation}
K(\eta,t)=\frac{G(\eta,t)}{\beta_{1}(t)}
\end{equation}
\begin{equation}
T=\int_{0}^{t}\frac{1}{\beta^{2}(s)}ds
\end{equation}

\begin{equation}
\frac{\partial K(\eta,T)}{\partial T}=D\frac{\partial^{2}K(\eta,T)}{\partial\eta^{2}}+\frac{Q(\eta,T)\beta^{2}(T)}{\beta_{1}}
\end{equation}

We now need to solve equation (80) to obtain a master equation for
transport of market particles in double auction limit order asset
markets. \citet{haberman_elementary_1987} provides a solution for
equations such as (80) using Greens Function Method (GFM). The solution
to (80) based on the above is given in (81):

\begin{equation}
 K(\eta,T)=\int_{0}^{T}\int_{-\infty}^{\infty}\frac{Q(\chi,\tau)\beta^{2}(\zeta)}{\sqrt{4\pi D(T-\zeta)}\beta_{1}}exp(-\frac{(\eta-\chi)^{2}}{4D(T-\zeta)})d\chi d\zeta
\end{equation}
\[
+\int_{-\infty}^{\infty}\frac{1}{\sqrt{4\pi DT}}G_{i}\omega(X)exp(-\frac{(\eta-\chi)^{2}}{4DT})d\chi
\]
Next we sequentially trace back the transformations done earlier,
in reverse order to get the solution below in (82). Here, $\zeta=\int_{0}^{\tau}\frac{1}{\beta^{2}(s)}ds$
and the initial condition $g(x,0)=G_{i}\omega(x)$.

\begin{equation}
 g(x,t)=\beta_{1}(t)\int_{0}^{t}\int_{-\infty}^{\infty}\frac{Q(\chi,\tau)}{\sqrt{4\pi D(T-\zeta)}}exp(\frac{(\frac{x}{\beta}-\int_{0}^{t}(\frac{\lambda_{2}}{\beta}+\frac{\lambda_{1}\phi_{2}}{\beta})ds-\chi)^{2}}{4D(T-\zeta)})d\chi d\tau
\end{equation}
\[
+\beta_{1}(t)\int_{-\infty}^{\infty}\frac{1}{\sqrt{4\pi DT}}G_{i}\omega(\chi)exp(-\frac{(\frac{x}{\beta}-\int_{0}^{t}(\frac{\lambda_{2}}{\beta}+\frac{\lambda_{1}\phi_{2}}{\beta})ds-\chi)^{2}}{4DT})d\chi
\]

\end{document}